\documentclass{aa}
\usepackage{natbib,twoopt}
\usepackage[breaklinks=true]{hyperref} 
\usepackage{tablefootnote}
\usepackage{threeparttable}
\usepackage{dcolumn}
\bibpunct{(}{)}{;}{a}{}{,} 
\makeatletter
\newcommandtwoopt{\citeads}[3][][]{\href{http://adsabs.harvard.edu/abs/#3}%
{\def\hyper@linkstart##1##2{}%
\let\hyper@linkend\@empty\citealp[#1][#2]{#3}}}
\newcommandtwoopt{\citepads}[3][][]{\href{http://adsabs.harvard.edu/abs/#3}%
{\def\hyper@linkstart##1##2{}%
\let\hyper@linkend\@empty\citep[#1][#2]{#3}}}
\newcommandtwoopt{\citetads}[3][][]{\href{http://adsabs.harvard.edu/abs/#3}%
{\def\hyper@linkstart##1##2{}%
\let\hyper@linkend\@empty\citet[#1][#2]{#3}}}
\newcommandtwoopt{\citeyearads}[3][][]%
{\href{http://adsabs.harvard.edu/abs/#3}
{\def\hyper@linkstart##1##2{}%
\let\hyper@linkend\@empty\citeyear[#1][#2]{#3}}}
\makeatother
\newcommand{\kms}{kms$^{-1}$}
\newcommand{\nod}{...}

\begin{document}
\newcolumntype{d}{D{.}{.}{2} }

\title{Stellar abundances and presolar grains trace the nucleosynthetic origin of molybdenum and ruthenium}

\titlerunning{Molybdenum and ruthenium}


\author{ C.\,J. Hansen\inst{1} \and A.\,C. Andersen\inst{2} \and N. Christlieb\inst{1}}
\authorrunning{Hansen, C. J. et al.}
\offprints{cjhansen, \email{@lsw.uni-heidelberg.de}}

\institute{Landessternwarte, ZAH, K\"onigstuhl 12, 69117  Heidelberg,
Germany
  \and 
Dark Cosmology centre, Niels Bohr Institute, University of Copenhagen, Juliane Maries Vej 30, DK-2100 Copenhagen, Denmark}

\date{today}

\abstract{This work presents a large consistent study of molybdenum (Mo) and ruthenium (Ru) abundances in the Milky Way. These two elements are important nucleosynthetic diagnostics. 
In our sample of 71 Galactic metal-poor field stars, we detect Ru and/or Mo in 51 of these (59 including upper limits). The sample consists of high-resolution, high signal-to-noise spectra covering both dwarfs and giants from [Fe/H] = $-0.63$ down to $-3.16$. Thus we provide information on the behaviour of Mo I and Ru I at higher and lower metallicity than is currently known. In this sample we find a wide spread in the Mo and Ru abundances, which is typical of heavy elements. We confirm earlier findings of Mo enhanced stars around [Fe/H] $= -1.5$ and add $\sim 15$ stars both dwarfs and giants with normal ($<0.3$\,dex) Mo and Ru abundances, as well as more than 15 Mo and Ru enhanced ($>0.5$\,dex) stars to the currently known stellar sample. This indicates that several formation processes, in addition to high entropy winds, can be responsible for the formation of elements like Mo and Ru. We trace the formation processes by comparing Mo and Ru to elements (Sr, Zr, Pd, Ag, Ba, and Eu) with known formation processes. Based on how tight the two elements correlate with each other, we are 
able to distinguish if they share a common formation process and how important this contribution is to the element abundance. We find clear indications of contributions from several different formation processes, namely the p-process, and the slow (s-), and rapid (r-) neutron-capture processes. From these correlations we find that Mo is a highly convolved element that receives contributions from both the s-process and the p-process and less from the main and weak r-processes, whereas Ru is mainly formed by the weak r-process as is silver. We also compare our absolute elemental stellar abundances to relative isotopic abundances of presolar grains extracted from meteorites. Their isotopic abundances can be directly linked to the formation process (e.g. r-only isotopes) providing a unique comparison between observationally derived abundances and the nuclear formation process. The comparison to abundances in presolar grains shows that the r-/s-process ratios from the presolar grains match the total elemental chemical 
composition derived from metal-poor halo stars with [Fe/H] around $-1.5$ to $-1.1$ dex. This indicates that both grains and stars around and above [Fe/H] $=-1.5$ are equally (well) mixed and therefore do not support a heterogeneous presolar nebula. An inhomogeneous interstellar medium (ISM) should only be expected at lower metallicities. Our data, combined with the abundance ratios of presolar grains, could indicate that the AGB yields are less efficiently mixed into stars than into presolar grains. Finally, 
we detect traces 
of s-process material at [Fe/H]$=-1.5$, indicating that this process is at work at this and probably at even lower metallicity.} 
 
\maketitle

\section{Introduction}
Nucleosynthetic processes are fundamental to the existence of stars, planets, and life. The neutron-capture processes can be traced through abundances of heavy elements (Z$>30$). Molybdenum and ruthenium are excellent trace elements because over time they can probe various enrichment scenarios, that contribute to the chemical evolution of our Galaxy in stars, meteorites, and planets. These two elements are created by three different processes, namely a rapid and slow neutron-capture process (the r- and s-processes, respectively), and a p-process. 

Owing to the lack of hyperfine structure, isotopic ratios of Mo and Ru cannot be determined by the stellar spectra covering a near-UV to visual range, and we can therefore only derive elemental stellar abundances. However, elemental abundances can be used to trace major contributions from various nucleosynthetic processes (as shown in \citealt{francois,roed10,hansen12}). The consistently analysed large sample allows us to make an indepth comparison of our data to abundances measured in presolar grains. The grain abundances directly probe which of the three processes --- r-, s-, or p --- contributed most to the stellar gas, as well as to the presolar grains at a given [Fe/H]. Furthermore, we can test whether there are changes as a function of time or metallicity in this contribution or, when phrased differently, if the dominating formation process varies. The formation processes that we try to trace may be associated with various sites, and here we only list a few (for details we refer to Fr\"ohlich et al. in prep.). 

The r-process seems to split into two, a main and a weak channel, where the main r-process could be linked to neutron-star mergers \citep[e.g.][]{freiburg,goriely}, while the weak r-process may come from neutrino-driven winds \citep[e.g.][]{arcones}, or electron-capturing supernovae (SNe) collapsing on O-Mg-Ne cores \citep{wan11}. A p-process contribution could also be expected from supernova winds that would facilitate for example a $\nu$p-process \citep{frohlich}. Finally, the parameter study of \citet{Fara} indicates that various r-process elements can be created in neutron-rich high entropy winds (HEW). Later in the Galactic chemical evolution s-process contributions from the main s-process associated with low-mass asymptotic giant branch (AGB) stars \citep{bisterzo,karakas, richard} and the weak s-process from, for instance, massive fast-rotating stars (see \citealt{pigna,frisch}) are expected to contribute to and possibly dominate the gas composition later on\footnote{According to \citet{cescutti}, these objects may also contribute with `primary' weak s-process yields very early in the history of our Galaxy.}. For reviews covering several of these formation processes, see \citet{chrisrev} and \citet{kaep11}.

Very few studies of both Ru and Mo abundances in metal-poor field stars have been performed. This is in part because their main absorption lines lie in the blue or near-UV part of the spectra, requiring very high-quality blue spectra for these studies.
Currently, our results compose the largest, most consistent sample of stars, for which Mo and Ru have been derived using 1D, local thermodynamic equilibrium based codes. \citet{ruth11} performed a large study of neutral and ionised Mo and Ru in both the near- and UV spectral ranges of a sample centered on [Fe/H] $\sim -1.5$. In our sample the stars span a slightly broader metallicity range and a wider Mo and Ru abundance range than previous studies.
Other studies have derived Mo I for single r-process enhanced stars \citep[e.g.][]{sneden03,ivans06,honda07,mashonkina10,roedMo10, siq13}. \citet{roederer12} also detected Mo II and Ru II in four extremely metal-poor stars from UV spectra. The sparse Mo detections of metal-poor halo field stars in the literature are also confirmed by Fr\"ohlich et al. (in prep), who carried out an extensive literature study. Furthermore, detections of Mo in globular clusters have been made by \citet{yong}, \citet{lai}, and \citet{roed11}, where the latter also report Ru abundances. Ruthenium was also studied in a metal-rich sample mainly consisting of Ba-stars \citep{allen}. Here we focus on metal-poor field stars and not on stars that may have been polluted by a binary or cluster environment.

Isotopic abundance ratios can be derived from presolar grains and other meteoritic samples. 
We present, for the first time, a comparison of Mo abundances in grains and stars spanning a broad range of metallicities.
While most of the material that went into making the solar system 
was thoroughly processed and mixed, thus losing isotopic heterogeneity and all memory of its origin, small quantities\footnote{1 to 100 parts per million in mass} of refractory presolar dust grains have been proven to be stardust with their original stellar isotopic signatures intact \citep{Hoppezinner00,Nittler03,Zinner05,Zinner98,OttHoppe07,Alexanderetal07,lodders2010}. These reflect the nucleosynthetic fingerprint of their stellar production site. 

The first presolar grains in meteorites were isolated by \citet{Lewisetal87} and 
identified as tiny diamonds. Later silicon carbide \citep[SiC;][]{Bernatowiczetal87}, graphite \citep{Amarietal90}, corundum \citep[Al$_{2}$O$_{3}$;][]{Hutcheon94,Nittleretal94}, 
silicon nitride \citep[Si$_{3}$N$_{4}$;][]{Russell95,Nittleretal95}, and 
spinel \citep[MgAl$_{2}$O$_{4}$;][]{Nittleretal94} were identified. 
Some of the SiC and graphite grains have been found to carry small 
inclusions of Ti-, Mo-, and Zr-carbides \citep{Bernatowiczetal91}.  Central to the identification of presolar grains is to determine the isotopic composition of the grain and/or some trace elements trapped in the grains. As a rule the isotopic composition of the grain or some of its inclusions deviates strongly from the normal solar system composition. The isotopic signatures of the grains contain information about the nucleosynthesis processes of the parent stars. Information on individual stars can be obtained by studying single grains by, for example, SIMS (secondary ion mass spectrometry) for the light-to-intermediate-mass elements, RIMS (resonance ionization mass spectrometry) for the heavy elements, and laser heating and gas mass spectrometry for He and Ne. TEM (transmission electron microscopy) and SEM (scanning electron microscopy) is used to study the crystal structure of the individual grains. 

The best characterised of all the presolar grains are SiC (6\,ppm) grains, and almost all of them originate in AGB stars. Graphite (less than 1\,ppm) was traditionally assumed to originate in supernovae, 
but more detailed measurements of $s$-process signatures indicate that most high-density graphite grains are more likely to originate in AGB stars \citep{Croatetal05}.  Since presolar grains seem to come from many stellar sources and to be relatively 
unprocessed, they also provide constraints on circumstellar dust production 
rates in the Galaxy. The relative abundances of the different types of presolar 
grains seem to indicate that AGB stars are the main dust producers in the Galaxy,
while supernovae only contribute a few percent \citep{Alexanderetal07}.

Isotopic abundances of Mo and Ru have been measured in presolar SiC and graphite grains. Since the amounts are quite small within the individual grains, it is on the edge of what is possible with the current techniques. In works by \citet{becker03}, \citet{lee}, and \citet{halliday03} it seems that the Mo isotopic abundances behave normal and seem to be well mixed into the matrix material of the meteorites, whereas \citet{Yin, Dauphas04,chen04}, on the other hand, find abundance anomalies for Mo (and Ru), indicating that the solar nebula was {\it not} homogeneous when the meteorites and the Earth\footnote{neither when the core formed nor later when the mantle formed} formed.

In this paper, we aim at finding evidence of any of these conclusions by studying elemental abundance in a large stellar sample and compare these to isotopic SiC grain abundances. 

Here we focus on the sample and stellar parameters in Sect. \ref{data}, and present the analysis in Sect. \ref{abanal}. The results will be described in Sect. \ref{res}, along with detailed abundance correlations from which clues to the nucleosynthetic formation processes that enriched the extremely metal-poor stars can be drawn. Finally, a detailed comparison of meteoritic isotopic abundances and elemental stellar abundances can be found in Sect. \ref{discus} along with our discussion, which will be followed by a short summary and conclusion in Sect. \ref{concl}. 

\section{Data and stellar parameters\label{data}}
This study is based on the analysis of the stellar sample described in \citet{hansen12}. The sample consists of 71 dwarfs and giants, observed with high-resolution spectrographs (UVES/VLT - \citet{Dekker}, and HIRES/Keck - \citet{vogt94}). The spectra are of $R \sim 40000 - \sim 50000$ in the blue. Owing to the excellent spectrum quality (a typical signal-to-noise ratio is S/N$>100$ at 3200\AA\,), abundances of Mo I and Ru I could be derived from blue and near-UV spectral lines. Details on data reduction can also be found in \citet{hansen12}.

\subsection{Continuum placement}
Continuum placement is a crucial part of the data analysis, especially when working in the blue spectral range. As noted in \citet[e.g.][]{ruth11,ruth13}, an over- or underestimation of the continuum level will lead to over- or underestimated Ru (or Mo) abundances. Therefore, we tested several fitting functions to ensure an accurate continuum placement. The echelle spectra were normalised in IRAF using the `continuum' package. We explored four fitting functions by varying the order of the fitted polynomial between one and seven and selecting the optimal fitting function, which may vary depending on the temperature, gravity, and metallicity of the stars. We found that the best functions were either fourth-order cubic splines or sixth-order Legendre polynomials. The function is fitted to the observed spectra, rejecting points that lie more than one sigma above or below the fitted function, and this process is repeated iteratively up to 30 times until the best match is found. Then the observed spectrum is divided by the best fit function (including rejected points).
This yields well-normalised spectra overall; however, each spectrum was visually inspected, and the procedure was repeated with different functions in case the result using for example a cubic spline was not satisfying. Before the abundances were derived, a region of $\pm20$\,{\AA} around the line of interest was zoomed in on and the continuum was re-evaluated and manually optimised, using the synthetic spectrum code MOOG \citep[][version 2010]{snedenphd}. By choosing a range of 40\,{\AA}, we avoid local biases that can be introduced by molecular bands, if for instance a region of only $\pm1$\,{\AA} was selected. However, with the 40\,{\AA} region we optimise the continuum, since this ensures that we include both absorption features and continuum pieces. Furthermore, it should be noted that the stellar sample was composed of stars without carbon or nitrogen enhancements. This means that there will be no strong molecular bands interfering with the continuum placement. 

To estimate how much this approach to setting the continuum level affects the derived abundances, every abundance was derived at least twice with a period of a couple of weeks between each abundance derivation. In this way, the derivations are largely independent, and the continuum placement, treatment of blends, and the overall fitting procedure can be assessed. For the most extreme cases, that is stars with high metallicity or metal-poor stars with slightly noisier spectra, the abundances derived in each of the analysis runs would differ by $\sim0.1$\,dex. In the spectra of the highest quality (high resolution and S/N), the abundances of each of the elements would typically agree within 0.01 to 0.03\,dex. Thus, the uncertainties related to fitting synthetic spectra spans a range of 0.01--0.1\,dex in the derived abundances (for details on abundance uncertainties see Sect. \ref{uncert}).

\subsection{Line blends and broadening}
Since the blue region is a crowded spectral region, line blending may be a concern, and it is therefore important to have a very complete description of lines in the vicinity of the line of interest. Thus, the line list was optimised as described below in Sect. \ref{abanal}. With this list at hand we can mimic the wing blends on each side of the Ru line, as well as reproduce the observed spectrum around Mo (see Figs. \ref{redMo} and \ref{spec}).
As mentioned above, no star with enhanced levels of N or C have been included in the sample. The effect of molecules (e.g. CN, CH, NH, etc.) was tested and found to be of the order of -0.02\,dex for Mo and -0.01\,dex for Ru. The test was carried out on the cold giant \object{BD+01 2916} where the effect of molecules is expected to be stronger than in warmer stars. From this the continuum placement is seen to be more important, influencing the Mo and Ru abundances more than molecular blends (owing to the sample selection).
This means that molecular features are very weak, and the line blending is therefore reduced to atomic lines. 

Finally, the broadening of the synthetic spectra was set in accordance with the instrumental resolution and fine-tuned so that the lines of moderate strength in the blue region would be matched. This was done by adjusting the full width half maximum in MOOG, and it was kept fixed when synthesising spectral lines.

\subsection{Stellar parameters\label{stelpar}}
The stellar parameters are taken from \citet{hansen12} and listed in Tables \ref{Mog} and \ref{Modw}. For the vast majority of the stars, these parameters have been determined using the following methods.
The effective temperatures are estimated using V$-$K-photometry (in a few cases we had to use excitation potential owing to the lack of K-magnitudes). The gravities are based on parallaxes when these are available (otherwise the ionisation equilibrium of Fe I and Fe II has been enforced), and the mictroturbulence is determined by requiring that all Fe I lines give the same abundance regardless of line strength (equivalent width). Finally, the metallicity ([Fe/H]) was derived from an average of Fe I and II abundances since these agree in general. For a few stars in which there is a large difference between Fe I and II, we have chosen the Fe II in order to minimise NLTE effects on the [Fe/H]. However, in some of the extremely metal-poor stars very few or no Fe II lines were present in spectra, and for these [Fe/H] has been based on Fe I abundance. For each set of stellar parameters (temperature, gravity, and metallicity) a MARCS model \citep{Gus08} was computed using the interpolation code from \citep{masseron}.
If the temperatures are derived from Fe lines, or if the gravity is set by enforcing ionisation equilibrium, the parameters have been labelled with an `a' in Tables \ref{Mog} and \ref{Modw}. In some cases the photometry or de-reddening were uncertain, while in others the parallaxes were inadequately determined, and the stellar parameters are tagged with a `c' to indicate they have been adjusted via excitation potentials and/or imposing ionisation equilibrium. Finally, stars with a special r-process pattern have been highlighted with a `b' in the same tables.

In general these stellar parameters agree with those recently derived for the Gaia-ESO benchmark stars \citep{paula}. For two of the three stars we have in common with their sample (\object{HD140283} and \object{HD22879}) the temperatures agree within 18\,K, the gravities within 0.07\,dex, and the [Fe/H] within 0.17\,dex. We also have \object{HD122563} in common with their work, but here the differences are larger (48\,K, 0.7\,dex, and 0.22\,dex).  However, when we compare our stellar parameters to those derived by \citet{Honda2004,honda07}, we find good agreement (18\,K, 0.2\,dex, 0.2\,dex) for \object{HD122563} and \object{HD88609}.

\section{Abundance analysis\label{abanal}}
\subsection{Atomic data}
The abundances were derived based on a line list downloaded from VALD \citep{vald} using `extract all'. The lines of interest as well as blends were cross checked with values from NIST\footnote{National Institute of Standards and Technology -- http://physics.nist.gov/asd} \citep{NIST} and similar studies from the literature. The log gf values we adopted for Mo I and Ru I (Table \ref{lines}) from VALD are the same as those given in \citet{sneden03,ivans06,ruth13}, and NIST. Additional line list information on molecules was taken from the database of Kurucz.

\begin{table}
\centering\caption{Atomic data from VALD and solar abundances from \citet{anders}.}
\begin{tabular}{ccccc}
\hline
\hline
Element & $\lambda$  & $\chi$& log gf & log $\epsilon_{\odot}$  \\
        & $[$\AA$]$  &   $[$eV$]$ &    &     $[$dex$]$  \\ 
\hline
Ru I & 3498.94   & 0.00 &  0.31  &  1.84          \\
Mo I & 3864.10   & 0.00 & -0.01  &  1.92  \\
Mo I & 5506.49   & 1.34 &  0.06  &  1.92 \\
\hline
\hline
\label{lines}
\end{tabular}
\end{table}
The lines we focussed on for this study are Mo I at 3864.1\,{\AA} and Ru I at 3498.9\,{\AA}. A lot of effort is currently going into assembling line lists for large surveys such as the Gaia-ESO Survey (GES). Thus we list the details on lines we have included and checked visibility in the solar spectra in Table \ref{gfs}. Especially the Mo I line at 5506\AA\, has been checked since it is included in the GES line list. 
\begin{table}
\centering
\caption{All persistent lines in the NIST database of Mo and Ru. Visible lines in the high-resolution solar NOAO atlas are marked with `*'.}
\begin{tabular}{ll}
\hline
\hline
 Mo I & Ru I \\
\hline
3133.59\,\AA\,&   3437.74\,\AA\,*\\
3158.17\,\AA\,&   3498.94\,\AA\,*\\
3170.34\,\AA\,&   3589.22\,\AA\,\\
3193.98\,\AA\,&   3593.03\,\AA\,\\
3208.84\,\AA\,&   3596.19\,\AA\,\\
3447.12\,\AA\,&   3726.93\,\AA\,\\
3798.25\,\AA\,*&  3728.03\,\AA\,*\\
3864.10\,\AA\,*&  3730.43\,\AA\,\\
3902.95\,\AA\,&   3798.90\,\AA\,*\\
4069.88\,\AA\,&   3799.35\,\AA\,*\\
4188.32\,\AA\,&   4199.89\,\AA\,\\
4411.70\,\AA\,&   \nod \\
5506.49\,\AA\,*&  \nod \\
5533.03\,\AA\,*&  \nod \\
5570.44\,\AA\,*&  \nod \\
\hline
\hline
\label{gfs}
\end{tabular}
\end{table}

\begin{figure}
\begin{center}
\includegraphics[width=0.49\textwidth]{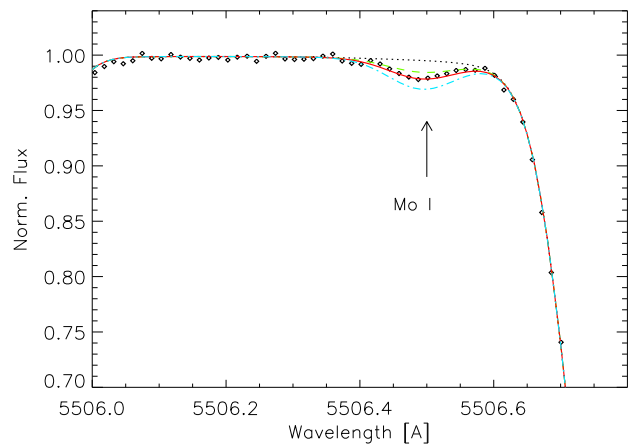}
\caption{Synthetic spectrum fit to the red Mo line at 5506\,\AA\, in the metal-rich dwarf HD113679 ([Fe/H]$=-0.63$). Spectra have been calculated with [Mo/Fe]$= -5., 0.05, 0.25$, and 0.45.}
\label{redMo}
\end{center}
\end{figure}

Out of these fairly strong (persistent\footnote{For example physics.nist.gov/PhysRefData/Handbook/Tables/\\rutheniumtable3.htm}) lines (see Table \ref{gfs}) only five Mo I and five Ru I lines (`*') were detectable. The bluest Mo I line is very close to the core of a hydrogen line and is therefore discarded, the three reddest Mo I lines are very weak, and the red 5506\,\AA\, line becomes too weak to detect below [Fe/H] = $-1$ in dwarfs. The red line will therefore only be used in the few `metal-rich' dwarf stars (see Fig. \ref{redMo}) in the sample\footnote{As seen from Table \ref{Modw}, the red 5506\,\AA\, line generally yields abundances for stars with [Fe/H] $> -1.0$\,dex unless the star is enhanced in Mo.}. This leaves us with one consistently useful Mo line, namely the 3864\,\AA\, Mo I line. The blending Fe 
lines were updated using atomic data from Kurucz's data base\footnote{http://www.cfa.harvard.edu/amp/ampdata/kurucz23/\\sekur.html}, since data neither from NIST nor VALD could properly reproduce the strong Fe-Sc-Cr blend blue of the 3864\,\AA\, Mo I line. 

\begin{figure}
\begin{center}
\includegraphics[width=0.49\textwidth]{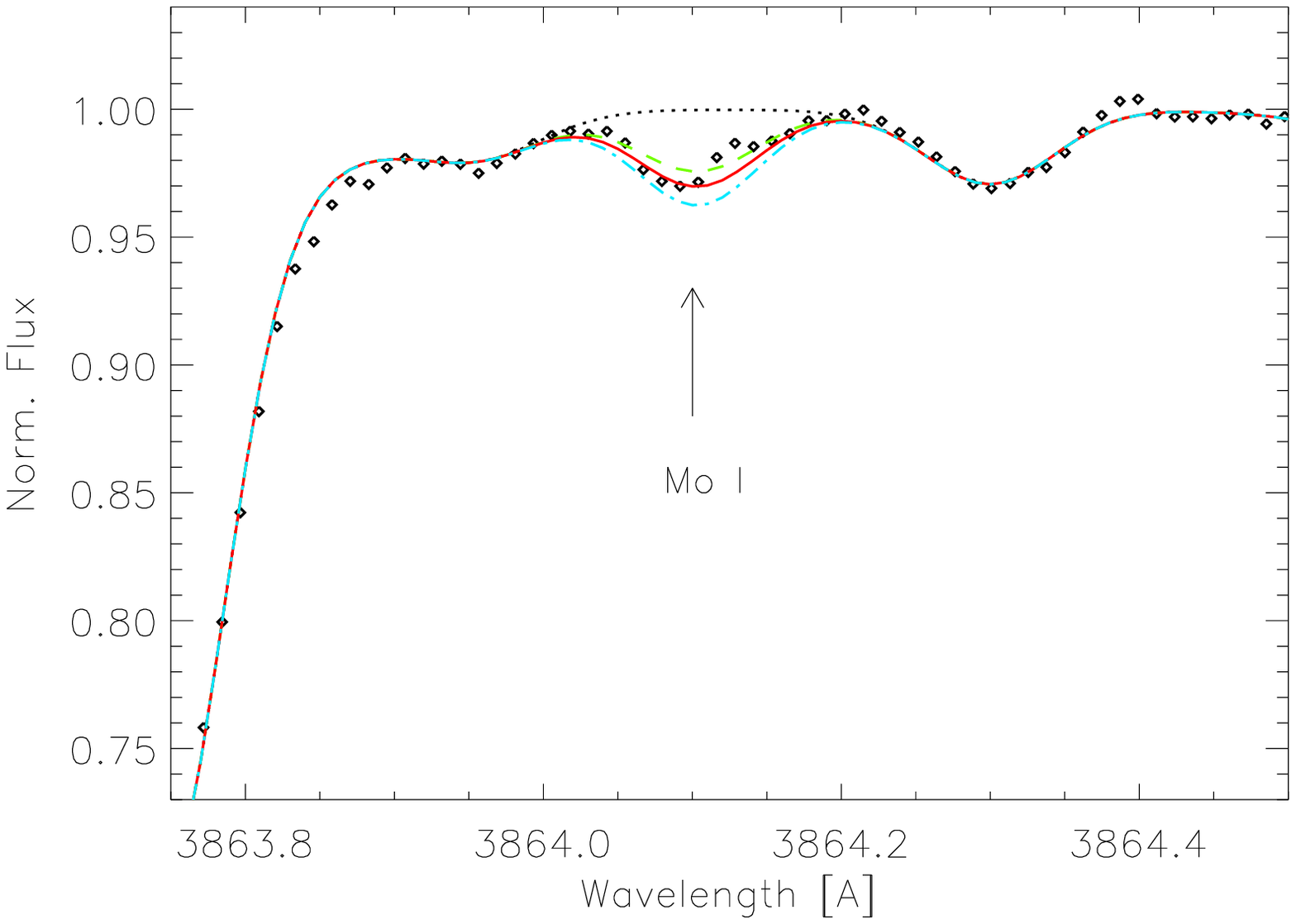}
\includegraphics[width=0.49\textwidth]{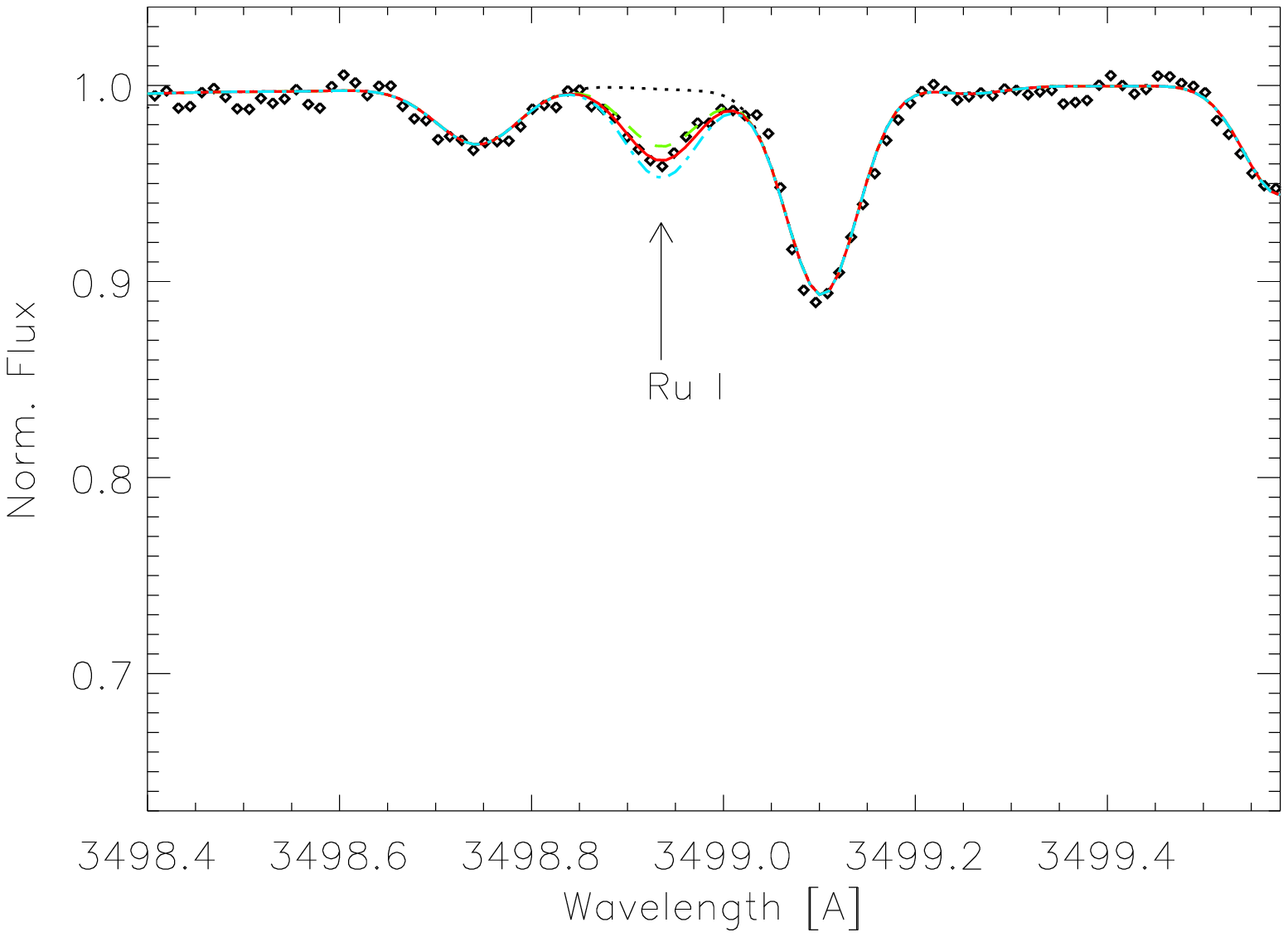}
\caption{Synthetic spectrum fit to the Mo line at 3864\AA\, (top) and the Ru line 3498\,\AA\, (bottom) in the metal-poor dwarf HD298986.  The signal-to-noise ratio is a little lower in this spectrum compared to that in Fig. \ref{hd1895}. Spectra have been calculated with [X/Fe]= -5., 0.5$\pm0.1$ or 0.66$\pm0.1$, where X is Mo or Ru, respectively. }
\label{spec}
\end{center}
\end{figure}

Similar findings are made for the Ru I lines. Here the two reddest lines fall in a strong hydrogen line, the 3728\,\AA\, line is heavily blended, and the blue most 3437\,\AA\, is blended and located in a NH band. The 3499\,\AA\, is the strongest and cleanest of all the detectable Ru lines.
The 3864\,\AA\, Mo line is blended (but it is the only usable line), while the Ru 3499\,\AA\, line is fairly strong and clean (with only a weak blend in the red wing -- see Fig. \ref{spec}). 
Therefore, this line yields the most trustworthy abundances. Further information can be found in Table \ref{lines}. Additionally, by only relying on these two lines, we can directly compare our derived abundances to those presented in \citet{ruth13}, which is currently the only other study detecting Mo and Ru in more than ten field stars in the Galaxy.
Figure \ref{spec} shows one of the most metal-poor warm dwarfs (\object{HD298986} with [Fe/H]$=-1.48$), for which both Mo and Ru has been derived to date\footnote{\object{HD188510} is 0.1\,dex more metal-poor, but the spectrum is of slightly lower quality.}.

\begin{figure}
\begin{center}
\includegraphics[width=0.49\textwidth]{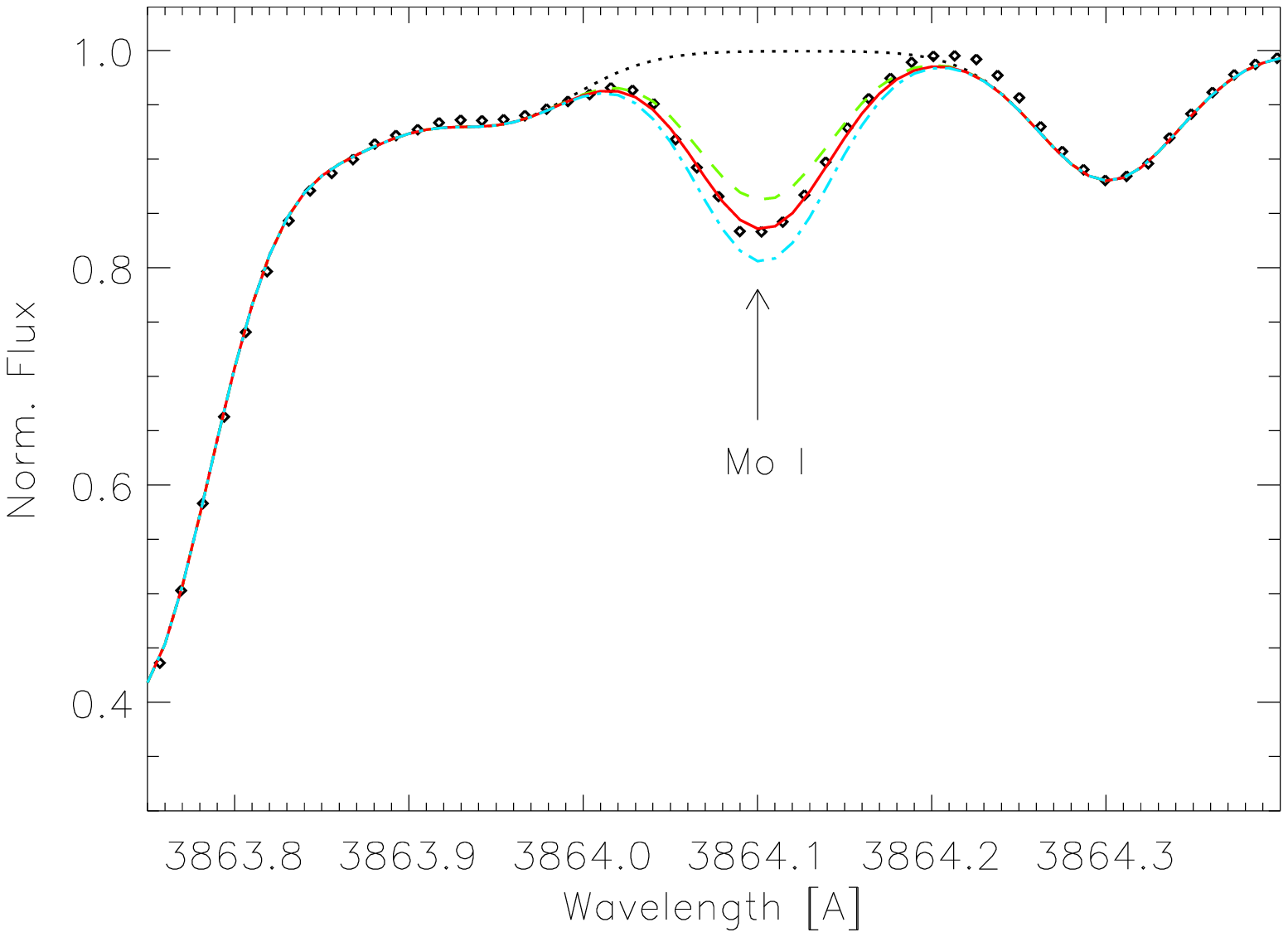}
\includegraphics[width=0.49\textwidth]{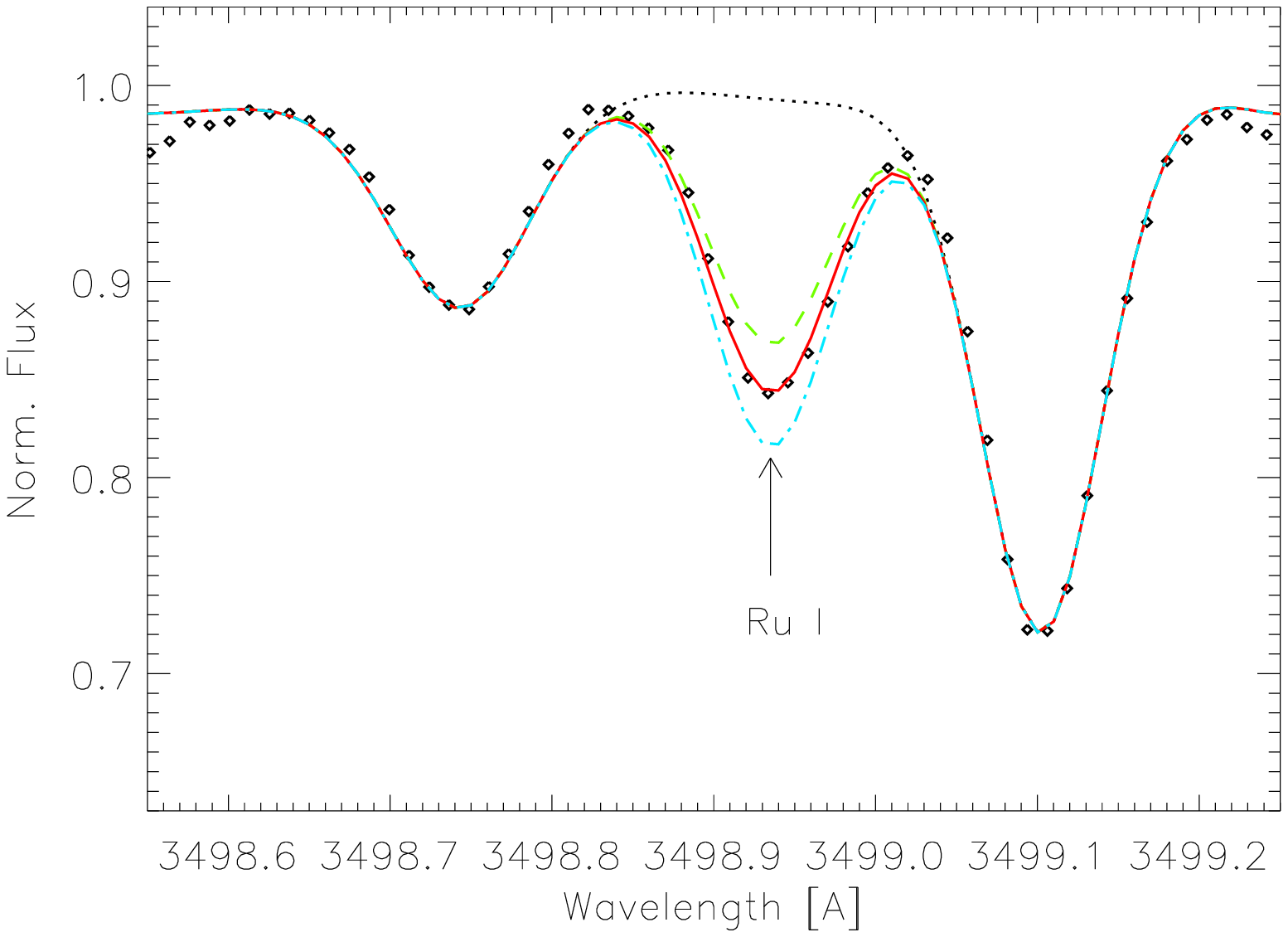}
\caption{High signal-to-noise spectrum of \object{HD189558} with synthetic spectra of [Mo/Fe] = $-5, 0.44, 0.54, 0.64$\,dex (top), and [Ru/Fe]$= -5, 0.49, 0.59, 0.69$\,dex (bottom) over-plotted. }
\label{hd1895}
\end{center}
\end{figure}

For comparison we have included a figure of \object{HD189558} that has an even higher signal-to-noise ratio than that of \object{HD298986}. A bit of noise $<1\%$ is seen in the Mo line, and slightly more around the Ru line in the spectrum of \object{HD298986}. In \object{HD189558} an almost perfect match between observed and synthesised spectrum is seen (Fig \ref{hd1895}). The spectra of both stars have been obtained with a resolution of $\sim 48000$, which for comparison is higher than the typical spectra ($R\sim 42000$) presented in \citet{ruth13}.

\subsection{Abundances of Mo and Ru\label{comparing}}
The stellar abundances were derived using MARCS 1D models \citep{Gus08}, and MOOG \citep{snedenphd} LTE synthetic spectrum code (version 2010). 
Our sample consists of 42 dwarfs and 29 giants, in which we detect either Mo and/or Ru in 30 (33) dwarfs and 22 (26) giants (including upper limits). We derive abundances only from neutral Mo and Ru lines. Both Mo and Ru have been detected down to [Fe/H] $=-3.0$ in \object{HD115444}, for which \citet{westin} did not have the needed spectral coverage to determine Ru. Upper limits for Mo and Ru are found for one of our most metal-poor stars (\object{HD126587}) with [Fe/H] $= -3.16$. In this sample we have 12 stars in common with the literature, namely \object{CS31082-001} for which \citet{siq13} derived [Mo/Fe] $= 0.9$ and [Ru/Fe] $=1.45$, which is in excellent agreement with our values of 0.94 and 1.44, respectively. The stellar parameters are also in good agreement (well within the combined errors), except for the microturbulence, which does not have any (or a very little) impact on these abundances since most of these are derived from weak lines. We also studied the two stars, \object{HD88609} and \object{HD122563}, for which we find a good agreement with the 
stellar parameters presented in \citet{honda07}, as mentioned in Sect. \ref{stelpar}. We find a maximum difference of 0.19\,dex between our Mo abundances and their results, a value that is within the combined uncertainties.
Three of the dwarf stars are \object{HD106038, HD160617}, and  \object{HD188510}, which we have in common with \citet{ruth13}. Our derived abundances are either in perfect agreement with those derived in \citet{ruth13} or agree within 0.1\,dex. The 0.1\,dex difference in Mo was found for \object{HD106038}, for which we derived 0.18\,dex different metallicities. We have two stars in common with \citet{ruth11}, namely \object{HD76932}, for which we find an excellent agreement between the stellar parameter derived, and the difference of 0.15\,dex in [Mo/Fe] probably stems from continuum placement. The second object is \object{HD140283}, for which neither we nor \citet{ruth11} could detect any Mo I nor Ru I. 

However, \citet{ruth11} studied UV spectra around $\lambda = 2000$\,{\AA} and could therefore derive abundances of Mo II and place an upper limit on Ru II in \object{HD140283}. Since the line lists for the UV region are often incomplete, owing to the lack of atomic data, it is common practice to empirically add oscillator strengths or even artificial lines 
to achieve a better match between the observed UV spectra and the synthesised ones. In \citet{ruth11}, they increase the log gf value of Mo II by 0.133 dex for five Mo II lines in the UV, to scale their theoretical values to that of the measured log gf for the Mo II line at 2082\,\AA. They note that the log gf values of Mo I and Mo II `appear to be on a consistent scale'\footnote{For comparison to Sr I and II we refer to \citet{hansen13} and Sect. \ref{res}}. Another study of Mo II in the UV is that of \citet{roederer12}, in which the 2871.51\,{\AA} line was used. In this study, they find a difference of up to $\sim 0.7$\,dex between the Mo I abundances and the Mo II upper limits. Since the studies in the UV often use adjusted log gf values, we would like to stress that possible NLTE effects might still affect both the Mo and Fe abundances, and that these effects are often seen to increase with increasing line strength (\citealt[for Fe and for Sr see][]{lindFe} \citealt{andr11,bergemann12} or \citealt{hansen13}). 

A larger study of Mo and Ru is that of \citet{ruth13}, in which Mo is detected in 20 stars (28 including upper limits). 
In this study the stellar parameters were determined from spectra (as ours labelled `a' or `c' in Tables \ref{Mog} and \ref{Modw}), and such stellar parameters can be off due to NLTE effects on Fe especially in stars with higher temperatures, lower metallicities, and gravities \citep{lindFe}. The NLTE corrections to the stellar parameters may in turn propagate into the stellar abundances and cause a larger change in abundance than the actual NLTE correction to the abundance itself. (This was shown to be the case for Sr II \citealt{hansen13}.)
\citet{ruth13} presents Mo I and Ru I abundances (between 0.4 -- 0.8\,dex). The two elements agree within 0.1\,dex, and show a low star-to-star abundance scatter. Therefore, they average the abundances of Mo and Ru and present this single value in order to better discuss the abundance enhancement with respect to other elements. 

Our results deviate from this picture (see Fig. \ref{MoRuFe}, as well as Tables \ref{Mog}, \ref{Modw}) because we find abundances in this larger sample that are not only enhanced but also only slightly above solar \citep[also found by][]{honda07,roederer12}; in other words, two groups with different enhancements of Mo and Ru are found. Here we find a large star-to-star abundance scatter as for most other heavy elements \citep[e.g.][]{francois,roed10,hansen12}. Furthermore, we show that Mo and Ru in several stars deviate by more than 0.1\,dex and that the elements are formed via different channels. Averaging these abundances therefore seems to erase important information, 
and we therefore choose to keep all heavy element abundances apart, compare them on a log $\epsilon$ abundance scale, and discuss their behaviour individually.

We derive Mo and Ru abundances that are only enhanced with a factor of two to three compared to the surrounding elements Zr, Pd, and Ag. However, we also find stars in the sample with Mo and Ru abundances that are of similar size as the Zr, Pd, and Ag abundances from \citet{hansen12}. Therefore, other sites than the high entropy winds \citep[HEW;][]{Fara} may be equally feasible (see Fr\"ohlich et al. 2014 in prep. for further comments on models and possible sites). 
There is also an overlap between this study and the four giants in \citet{roederer12}. Most of the stellar parameters are in good agreement with those presented here (adopted from \citealt{hansen12}), and the derived Mo and Ru abundances agree within 0.1\,dex. A slight disagreement in the determined stellar parameters of \object{HD108317} causes a slightly larger difference in the derived abundances of up to 0.24\,dex. This difference is, however, well within the combined abundance uncertainties. 

To obtain the purest trace of the nucleosynthetic formation processes, absolute (log $\epsilon$) abundances have been compared to those from \citet{hansen12}. The abundances have been obtained in exactly the same way using the same atmospheric models, methods for determining stellar parameters, as well as the same synthetic spectrum code (though here we used a more recent version). This allows a direct and unbiased comparison to those abundances (of Sr, Zr, Pd, Ag, Ba, and Eu). The derived Mo and Ru abundances have been plotted vs [Fe/H] in Fig. \ref{MoRuFe}. The [Mo/Fe] and [Ru/Fe] have been calculated using the Mo$_{\odot}=1.92$ and Ru$_{\odot}=1.84$ from \citet[][see our Table \ref{lines}]{anders}, which will facilitate a direct comparison to the meteoritic abundances discussed in Sect. \ref{discus}.

\subsection{Uncertainties\label{uncert}}
The abundance uncertainties resulting from uncertainties in stellar parameters, continuum placement in the near-UV, and fitting technique have been determined in great detail for a sample dwarf and giant star. The two selected stars are listed in Table \ref{table:uncert}. They were chosen since they have a good, clean spectral region around the lines. This allows an easier assessment of the stellar parameter's impact on the abundances. Furthermore, these stars have stellar parameters that are representative of the dwarfs and giants in this sample. The general uncertainties on the stellar parameters adopted from \citet{hansen13} are for temperature/gravity/[Fe/H]/$\xi$: $\pm$100\,K$/0.2/0.15/0.15$\,\kms, and fitting related errors span a factor of 10 when associated with fitting technique (best case $\pm0.01$\,dex) and continuum placement ($\pm 0.1$\,dex). The uncertainty related to the synthetic spectrum fitting and continuum placement also includes an assessment of the line blending. The strength of the blending lines were varied to test the impact on the derived abundances. The total propagated error is given in Table \ref{table:uncert}, where each term has been added in quadrature. 
\begin{table*}
\caption{Abundance uncertainties given on a $\log \epsilon-$scale for a dwarf (\object{HD188510}) and giant (\object{BD+54 1323}) star.}
\begin{tabular}{lcccc}
\hline
\hline
  & \multicolumn{2}{c}{Dwarf} & \multicolumn{2}{c}{Giant} \\
Parameter & Mo & Ru & Mo & Ru \\
\hline
T $\pm100$K       & 0.11 &  0.1  & 0.08 & 0.13 \\
$\log$g $\pm0.2$  & 0.02 &  0.01 & 0.03 & 0.01 \\
$[$Fe$/$H$] \pm0.15$  & 0.01 &  0.04 & 0.02 & 0.01 \\
$\xi \pm0.15$     & 0.01 &  0.03 & 0.03 & 0.01 \\
Fitting& 0.01/0.1 & 0.01/0.1 & 0.01/0.1 & 0.01/0.1 \\ 
\hline
Total $\pm$:& 0.11(0.15)& 0.12(0.15) & 0.11(0.14) & 0.14(0.16) \\
\hline
\hline
\label{table:uncert}
\end{tabular}
\end{table*}

As the continuum in the near-UV is often challenging to place, particularly in stars with [Fe/H]$>-2.5$, we adopt the larger uncertainty given in parentheses in all the figures. The uncertainty is $\sim 0.15$\,dex on average.

Generally, the spectra could be well fitted; however, in a few cases where the spectral quality was a bit lower, or where the metallicity of the star was high, the line fit would become too poor, and we could therefore only place upper limits. More than four observational data points had to be well matched for the fit to be a detection, otherwise it would lead to upper limits. We have a few such cases in our sample, where the metallicity is above $-1$\,dex (causing line blending), and the spectral quality is lower, namely \object{HD175179, G005-040, HD105004, CD-45 3283}. The two former stars also have upper limits on Sr and Y as shown in \citet{hansen12}. The molybdenum falls at the end of the blue spectrum (when using the UVES/VLT 346nm setting), and this means that the spectra generally will be more noisy around the Mo line than around the Ru line. For the last two stars, as well as \object{HD111980}, the spectra end less than 2-3{\AA} from the Mo line, either due to radial velocity and/or spectrum 
quality, and we can therefore not place the continuum properly in this noisy region. Thus we present upper limits for these stars.

\section{Results\label{res}}
\begin{figure}
\begin{center}
\includegraphics[width=0.52\textwidth]{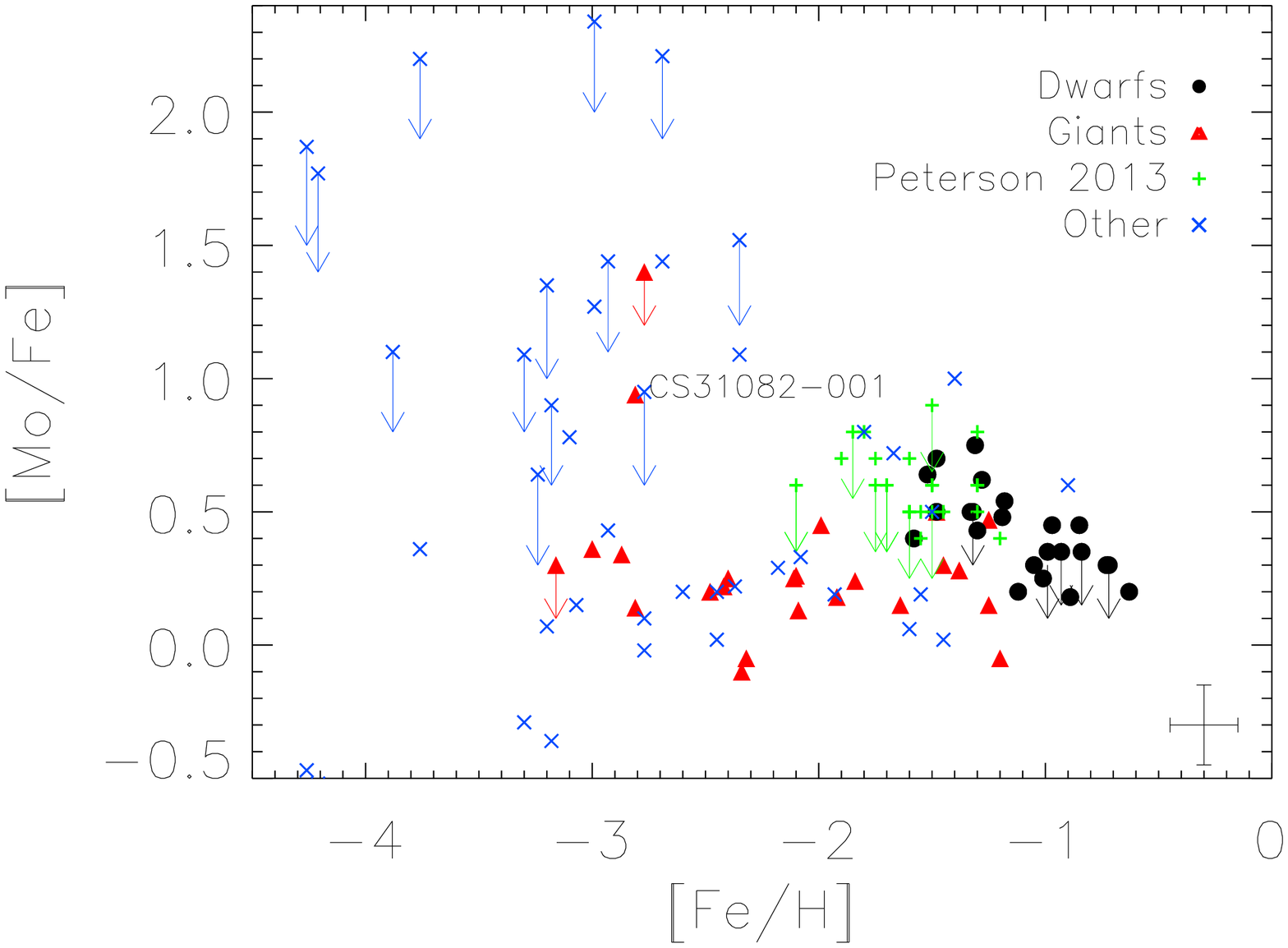}
\includegraphics[width=0.52\textwidth]{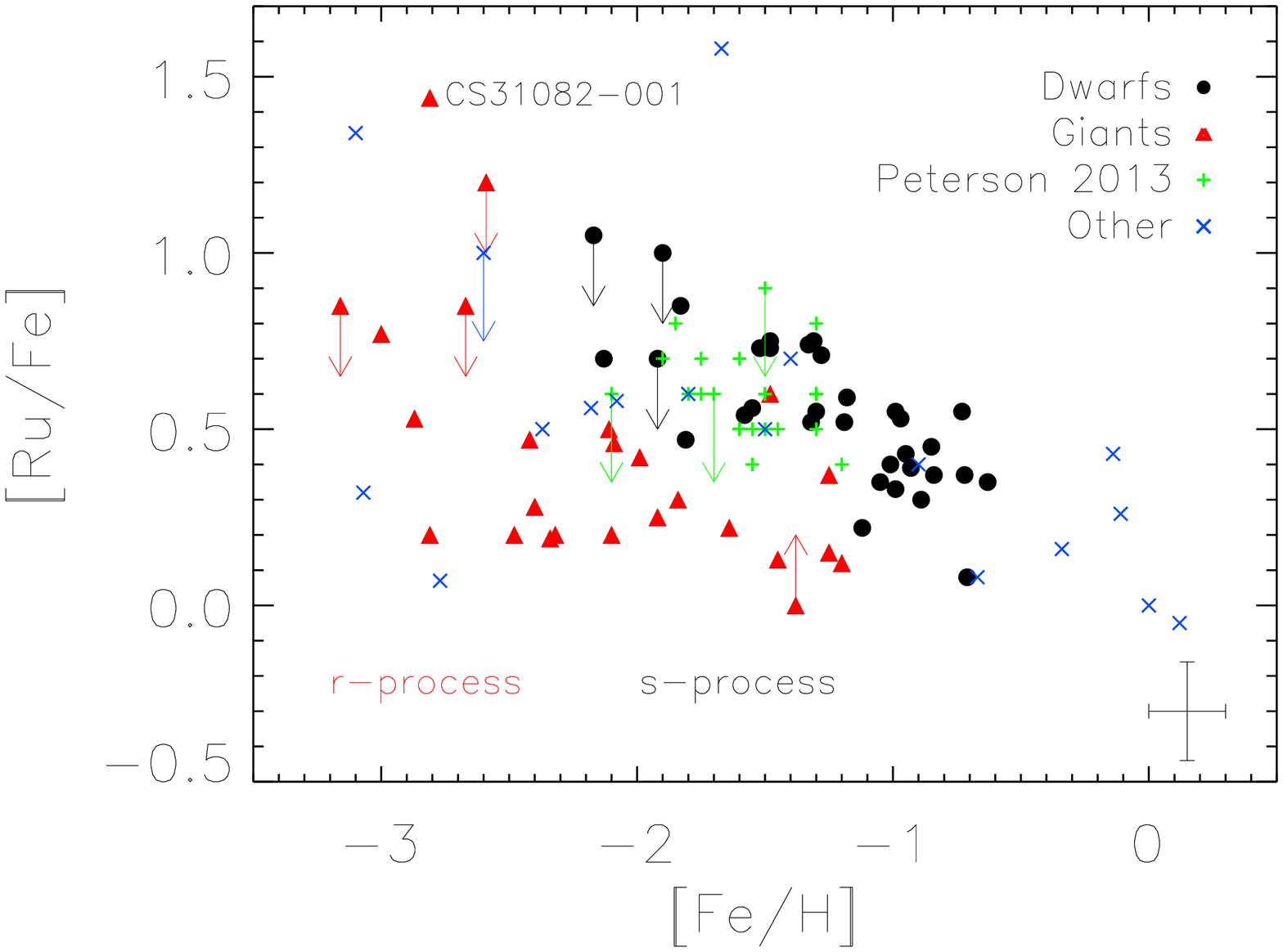}
\caption{Top: [Mo/Fe] as a function of metallicity. Dwarf stars are shown as black filled circles, giants as filled red triangle, while comparison stars from \citet{ruth13} are shown as green pluses. Bottom: [Ru/Fe] as function of [Fe/H] using the same symbols as just described. `Others' is a sample comprised data from: \citet{sneden03,ivans06,allen,honda07,roed10,roedMo10,ruth11,roederer12,chrjohn,ianUPL2014}.\label{MoRuFe}}
\end{center}
\end{figure}

In total we derived Mo abundances for 47 stars and Ru abundances for 58 stars, with clear detections down to $-3.0$ in [Fe/H].
The results are listed in Tables \ref{Mog} and \ref{Modw} and the abundances relative to iron are shown in Fig. \ref{MoRuFe}.  As for many of the heavy elements, Mo and Ru also show a large star-to-star scatter. For the Mo and Ru enhanced stars, we find a good agreement and overlap with the metal-poor sample from \citet{ruth13}.

Figure \ref{MoRuFe} shows our [Mo/Fe] and [Ru/Fe] compared to other samples as a function of [Fe/H]. The comparison samples are comprised of \citet{ruth13} study, and all the other studies of less than ten stars have been grouped together. Here we note that only stars that were labelled as not being Ba stars from the sample of \citet{allen} have been included. This is a large study of 39 stars (33 Ba stars) for which normal (extreme) Ru abundances have been derived for stars with metallicities above $\sim -0.5$\,dex. \citet{ruth11} is incorporated in the `Others' sample, as is the extremely Ru enhanced bulge giant star from \citet{chrjohn}, and the upper limits on Mo from \citet{ianUPL2014}.
For the 12 stars we have in common with the literature we have shown both our own and previous measurements (using different symbols), and for eight of these duplicates the agreement is so good that the symbols touch or overlap. For the remaining stars a difference of $\sim0.2$\,dex in either Mo or Fe lead to larger abundance differences (see Sects. \ref{data} and \ref{comparing}). Stars with normal and enhanced Ru and Mo abundances are found both in dwarfs and giants, regardless of which method was applied to determine stellar parameters. We note that the exact abundances of dwarfs and giants could change due to NLTE effects on stellar parameters, as well as on Mo and Ru abundances\footnote{A similar study was carried out by \citet{hansen13} and \citet{bergemann12} on Sr I and II, where large NLTE corrections were found for Sr I, while the largest impact on Sr II stems from the NLTE corrections to the stellar parameters.}. There is a slight gap around [Fe/H]$=-2.5$ in [Mo/Fe]. More Mo detections around this [Fe/H] would be 
needed to see whether the enhanced trend in the dwarfs would continue towards lower metallicities, or if they would follow the 
chemical normal pattern. However, this could be an observational bias that is not easily overcome (see below).

Since the stellar abundances may be biased by the 1D, LTE assumptions, we tried to assess the bias this assumption may lead to by separating dwarfs and giants, since they most likely will be affected differently by these assumptions. Thus, an offset in trend found between the dwarfs and giants can be used as a possible probe of the size and impact of these assumptions compared to 3D, NLTE abundances. Unfortunately, no model atom exits for Mo or Ru, which is why we cannot correct our LTE abundances. 

A difference in the abundance behaviour between dwarfs and giants can also be induced by a difference in metallicity. Most of the Mo and Ru absorption lines disappear in the extremely metal-poor dwarfs, while they remain detectable in the giants. This means that the giants can probe the abundance behaviour at lower metallicities, and will therefore (despite possible deviations from LTE and 1D) be better tracers of the earliest formation processes (r- and possibly p-processes). The dwarfs, on the other hand, mainly show Mo and Ru lines above [Fe/H]$\sim -2$, and will therefore trace later epochs of the chemical evolution of the Galaxy, which have been enriched (and dominated?) by the s-process(es) and smaller p-process contributions (see Fig. \ref{MoRuFe}). At high metallicities ([Fe/H]$> -1.1$), we can expect to see both s- and p-process contributions from SN type Ia. The exact yields are model sensitive \citep[see][for details]{trav11}. 
This will be the overall trend we can expect to see from the elemental abundances, whereas the isotopic abundances of Mo and Ru will each have a direct link to the individual formation process(es) that create that particular isotope. For guidance these have been listed in Table \ref{isoabu}.

\begin{table*}
\caption{Stellar abundances of Mo and Ru for giant stars. }
\begin{tabular}{lccccdd}
\hline
\hline
Giants          & T  &  log g & $\xi $ & [Fe/H]  & \centering\mathrm{[Mo/Fe]} & \mathrm{[Ru/Fe]}  \\      
                & [K] &   &   [km/s] &   & &   \\      
\hline
BD-01 2916  	& 4480$^a$  & 1.20$^a$  &  2.4 &  $-1.99$  &   0.45    & 0.42   \\
BD+8 2856	& 4600$^a$  & 0.80$^a$  &  2.0 &  $-2.09$  &   0.13^{cp}  & 0.46^+ \\
BD+30 2611	& 4238      & 0.50$^a$  &  1.7 &  $-1.20$  &  -0.05    & 0.12   \\
BD+42 621	& 4725$^a$  & 1.50$^a$  &  1.7 &  $-2.48$  &   0.20    & 0.20   \\ 
BD+54 1323	& 5213      & 2.01$^c$  &  1.5 &  $-1.64$  &   0.15    & 0.22   \\ 
CS22890-024	& 5400      & 2.65$^a$  &  1.7 &  $-2.77$  &  <1.4     &  \nod     \\ 
CS29512-073	& 5000$^a$  & 1.85$^a$  &  1.1 &  $-2.67$  &    \nod      &  <0.85 \\ 
CS30312-100	& 5200      & 2.35$^a$  &  1.4 &  $-2.62$  &    \nod      &  \nod     \\ 
CS30312-059	& 5021      & 1.90$^a$  &  1.5 &  $-3.06$  &    \nod      &  \nod     \\ 
CS31082-001$^b$	& 4925      & 1.51$^a$  &  1.4 &  $-2.81$  &    0.94   &  1.44  \\
HD74462         & 4590      & 1.84$^c$  &  1.1 &  $-1.48$  &    0.50   &  0.60  \\ 
HD83212         & 4530      & 1.21$^c$  &  1.8 &  $-1.25$  &    0.15^+ &  0.15^+\\ 
HD88609$^b$	& 4568      & 1.01$^c$  &  1.9 &  $-2.87$  &    0.34   &  0.52  \\
HD108317	& 5360      & 2.76      &  1.2 &  $-2.11$  &    0.20   &  0.50  \\ 
HD110184	& 4450$^a$  & 0.50$^c$  &  2.1 &  $-2.40$  &    0.25   &  0.28  \\ 
HD115444$^b$	& 4785      & 1.50$^c$  &  2.1 &  $-3.00$  &    0.36   &  0.77  \\
HD122563$^b$	& 4560$^a$  & 0.90$^a$  &  1.8 &  $-2.81$  &    0.14   &  0.20  \\ 
HD122956	& 4700      & 1.51      &  1.2 &  $-1.45$  &    0.30   &  0.13  \\
HD126238	& 4900      & 1.80      &  1.5 &  $-1.92$  &    0.18   &  0.25  \\
HD126587	& 4700$^a$  & 1.05$^c$  &  1.7 &  $-3.16$  &   <0.3    &  <0.85 \\ 
HD128279	& 5200$^a$  & 2.20$^a$  &  1.3 &  $-2.34$  &   -0.10   &  0.19  \\ 
HD165195	& 4200$^c$  & 0.90$^c$  &  2.1 &  $-2.10$  &    0.26   &  0.20  \\ 
HD166161$^b$	& 5250$^a$  & 2.15$^c$  &  1.9 &  $-1.25$  &    0.47   &  0.37  \\
HD175305	& 5100      & 2.70      &  1.2 &  $-1.38$  &    0.28   &  >0.0* \\
HD186478	& 4730      & 1.50$^c$  &  1.8 &  $-2.42$  &    0.22   &  0.47  \\
HD204543	& 4700      & 0.80$^a$  &  2.0 &  $-1.84$  &    0.24   &  0.30  \\
HE0315+0000	& 5200      & 2.40$^a$  &  1.6 &  $-2.59$  &    \nod   &  <1.2  \\ 
HE0442-1234	& 4530      & 0.65$^a$  &  1.8 &  $-2.32$  &   -0.05   & 0.20   \\
HE1219-0312	& 4600      & 1.05$^a$  &  1.4 &  $-3.21$  &    \nod   & \nod      \\
\hline
\hline
\label{Mog}
\end{tabular}
\begin{tablenotes}
            \item [1] $^*$ Spike in spectral line.
	    \item [2] $^+$ Larger uncertainty in the abundance due to blends.
	    \item [3] $^{cp}$ Continuum placement uncertain.
	    \item [4] $^a,^b,^c$: a) T is derived from excitation potentials, $\log$ g from ionisation equilibrium, b) the star has a special r-process pattern, c) uncertain colour, dereddening, or parallax lead to adjustment according to a).
        \end{tablenotes}
\end{table*}

Table \ref{Modw} shows Mo abundances derived from the blue 3864\,{\AA} and the red 5506\,{\AA} line. Generally these two lines yield exactly the same abundance (or values within 0.02\,dex or 0.14\,dex when an upper limit is derived). This good agreement indicates that the abundances derived from the blue Mo line have not been biased by, for instance, continuum placement or line blending. We choose to list the blue and red Mo abundance in different columns to allow a direct comparison between our blue line values and other metal-poor studies, as well as future studies of the red line carried out as part of surveys such as GES. Based on the detections of Mo from the red 5506\,{\AA} line, there seems to be a cut in stellar parameters for which Mo can be detected. These indicate that Mo can in general be detected in cool and warm dwarfs down to [Fe/H]$\sim -0.7$, upper limits down to $-1$\,dex, and if the star is enhanced or cool ($<5700$\,K), the line is useful down to [Fe/H]$\sim -1.3$. The values presented in our 
figures are the abundances from the blue 3864\,{\AA} Mo I line. 

\begin{table*}
\caption{Stellar abundances of Mo and Ru for dwarf stars.}
\begin{tabular}{lccccddd}
\hline
\hline
Dwarfs	        & T & log g & $\xi $ &  [Fe/H]  & \centering\mathrm{[Mo/Fe]} &\centering\mathrm{[Mo/Fe]} & \mathrm{[Ru/Fe]}  \\
     & [K] &  & [km/s] &  & \centering\mathrm{(3864A)} &\centering\mathrm{(5506A)} &   \\
\hline
BD+09 2190       & 6450$^a$  & 4.00      &  1.5 &  $-2.60$  &  \nod   & \nod 	&   \nod    \\ 
BD-13 3442       & 6450      & 4.20$^a$  & 1.5  &  $-2.56$  &  \nod   & \nod	&   \nod    \\  
CD-30 18140      & 6340      & 4.13      & 1.0  &  $ -1.92$ &  \nod   & \nod	&   <0.7  \\  
CD-33 3337      & 5952      & 3.95      & 1.4  &  $-1.55$  &   \nod   & \nod    &   0.56  \\  
CD-45 3283      & 5657$^c$  & 4.97      & 0.8  &  $-0.99$  &   0.31   & <0.25   &   0.55  \\
CD-57 1633      & 5907      & 4.26      & 1.1  &  $-1.01$  &   \nod   & \nod 	&   0.40^+\\  
HD3567          & 6035      & 4.08      & 1.5  &  $-1.33$  &    0.50  & \nod	&   0.74  \\  
HD19445         & 5982      & 4.38      & 1.4  &  $-2.13$  &   \nod   & \nod 	&   0.70  \\  
HD22879         & 5792      & 4.29      & 1.2  &  $-0.95$  &   \nod   & \nod	&   0.43  \\  
HD25704         & 5700      & 4.18      & 1.0  &  $-1.12$  &    0.20  & \nod 	&   0.22  \\  
HD63077         & 5629      & 4.15      & 0.9  &  $-1.05$  &    0.30  & \nod	&   0.35  \\  
HD63598         & 5680      & 4.16      & 0.9  &  $-0.99$  &   \nod   & <0.0	&   0.33  \\  
HD76932         & 5905      & 4.08      & 1.3  &  $-0.97$  &    0.45  & \nod	&   0.53  \\  
HD103723        & 6128      & 4.28      & 1.5  &  $-0.85$  &    0.45  & \nod 	&   0.45  \\  
HD105004        & 5900$^a$  & 4.30$^c$  & 1.1  &  $-0.84$  &   <0.35  & \nod	&   0.37  \\
HD106038$^b$    & 5950      & 4.33      & 1.1  &  $-1.48$  &    0.70  & \nod	&   0.75  \\
HD111980$^b$    & 5653      & 3.90      & 1.2  &  $-1.31$  &    0.51  & <0.5    &   0.52  \\ 
HD113679        & 5759      & 4.04      & 0.9  &  $-0.63$  &    0.22  &  0.2    &   0.35  \\ 
HD116064        & 5999      & 4.33      & 1.5  &  $-2.19$  &    \nod  & \nod 	&   <1.05 \\ 
HD120559        & 5411      & 4.75      & 0.7  &  $-1.33$  &    0.75  & \nod 	&   0.75  \\ 
HD121004        & 5711      & 4.46      & 0.7  &  $-0.73$  &    \nod  & 0.3 	&   0.55  \\ 
HD122196        & 6048      & 3.89      & 1.2  &  $-1.81$  &    \nod  & \nod 	&   0.47  \\ 
HD126681$^b$    &  5532     & 4.58      & 0.6  &  $-1.28$  &    0.62  &  0.6	&   0.71  \\ 
HD132475        & 5838      & 3.90      & 1.5  &  $-1.52$  &    0.64  & \nod 	&   0.73  \\ 
HD140283        & 5738      & 3.73      & 1.3  &  $-2.58$  &    \nod  & \nod 	&   \nod  \\ 
HD160617        & 6028      & 3.79      & 1.3  &  $-1.83$  &    \nod  & \nod 	&   0.85  \\ 
HD166913        & 6155      & 4.07      & 1.5  &  $-1.30$  &    0.43  & \nod 	&   0.55  \\ 
HD175179        & 5758      & 4.16      & 0.9  &  $-0.72$  &    0.4^{cp} &  <0.3&   0.37  \\ 
HD188510        & 5536      & 4.63      & 1.0  &  $-1.58$  &    0.40  & \nod 	&   0.54  \\ 
HD189558        & 5712      & 3.79      & 1.2  &  $-1.18$  &    0.54  &  <0.68	&   0.59  \\ 
HD195633        & 6005      & 3.86      & 1.4  &  $-0.71$  &    \nod  & \nod	&   0.08  \\ 
HD205650        & 5842      & 4.49      & 0.9  &  $-1.19$  &    0.48  & \nod	&   0.52  \\ 
HD213657        & 6208      & 3.78      & 1.2  &  $-2.01$  &    \nod  & \nod 	&   \nod  \\ 
HD298986        & 6144      & 4.18      & 1.4  &  $-1.48$  &    0.50  & \nod 	&   0.66  \\ 
G005-040        & 5766      & 4.23$^a$  & 0.8  &  $-0.93$  &    0.30  & <0.35  &  0.39   \\
G013-009        & 6416      & 3.95      & 1.4  &  $-2.27$  &    \nod  & \nod	&  \nod    \\ 
G020-024        & 6482      & 4.47      & 1.5  &  $-1.89$  &    \nod  & \nod 	&  <1.0   \\  
G064-012        & 6459      & 4.31$^c$  & 1.5  &  $-3.10$  &    \nod  & \nod 	& \nod      \\  
G064-037        & 6494      & 3.82$^c$  & 1.4  &  $-3.17$  &    \nod  & \nod 	& \nod      \\  
G088-032        & 6327      & 3.65      & 1.5  &  $-2.50$  &    \nod  & \nod 	& \nod      \\ 
G088-040        & 5929      & 4.14      & 1.4  &  $-0.90$  &    0.18  & \nod 	&  0.30   \\  
G183-011        & 6309      & 3.97      & 1.0  &  $-2.12$  &    \nod  &  \nod	& \nod     \\
\hline
\hline
\label{Modw}
\end{tabular}
\begin{tablenotes}
	    \item [1] $^+$ Larger uncertainty in the abundance due to blends.
	    \item [2] $^{cp}$ Continuum placement uncertain.
	    \item [3] $^a,^b,^c$: a) T is derived from excitation potentials, $\log$ g from ionisation equilibrium, b) the star has a special r-process pattern, c) uncertain colour, dereddening, or parallax lead to adjustment according to a).
        \end{tablenotes}
\end{table*}
\begin{table}
\caption{Formation processes of Mo and Ru isotopes \citep{Dauphas04}. `P' indicates that the isotope is created only by a p-process, similarly for `s' and `r', while `r+s' means that the isotope can be created by both r- and s-processes.}
\begin{tabular}{cccccccc}
\hline
\hline
Element&&\multicolumn{5}{c}{Isotope} &\\
\hline
Mo & 92 & 94 & 95 & 96 & 97 & 98 & 100 \\
Ru & 96 & 98 & 99 & 100 & 101 & 102 & 104 \\
\hline
Process & p & p & r + s & s & r + s& r + s & r\\
\hline
\hline
\label{isoabu}
\end{tabular}
\end{table}

\subsection{Finding correlations\label{correl}}
To trace the nucleosynthetic origin of Mo and Ru we compare these elements to others with known formation processes. It has been suggested that a weak r-process creates elements in the range $40< Z <50$, and \citet{hansen12} showed that Ag, and to some extent Pd, were created by this second weak r-process. These observations show not only that there was one universal r-process creating all r-process dominated elements heavier than Zn, but also that a weak r-process interfered with this picture of the trans-iron elements. Here we explore if this process also contributes to the formation of Mo and Ru.

We start by comparing the stellar derived elemental abundances of Mo and Ru to abundances of Sr, Zr, Pd, Ag, Ba, and Eu derived by \citet{hansen12}. The abundances of these six elements have been derived using the same method and codes for the analysis as the ones applied here. Thus, it is a very homogeneous analysis, which allows a direct comparison between the previously published abundances and the ones derived here. 

\begin{figure}
\begin{center}
\includegraphics[width=0.5\textwidth]{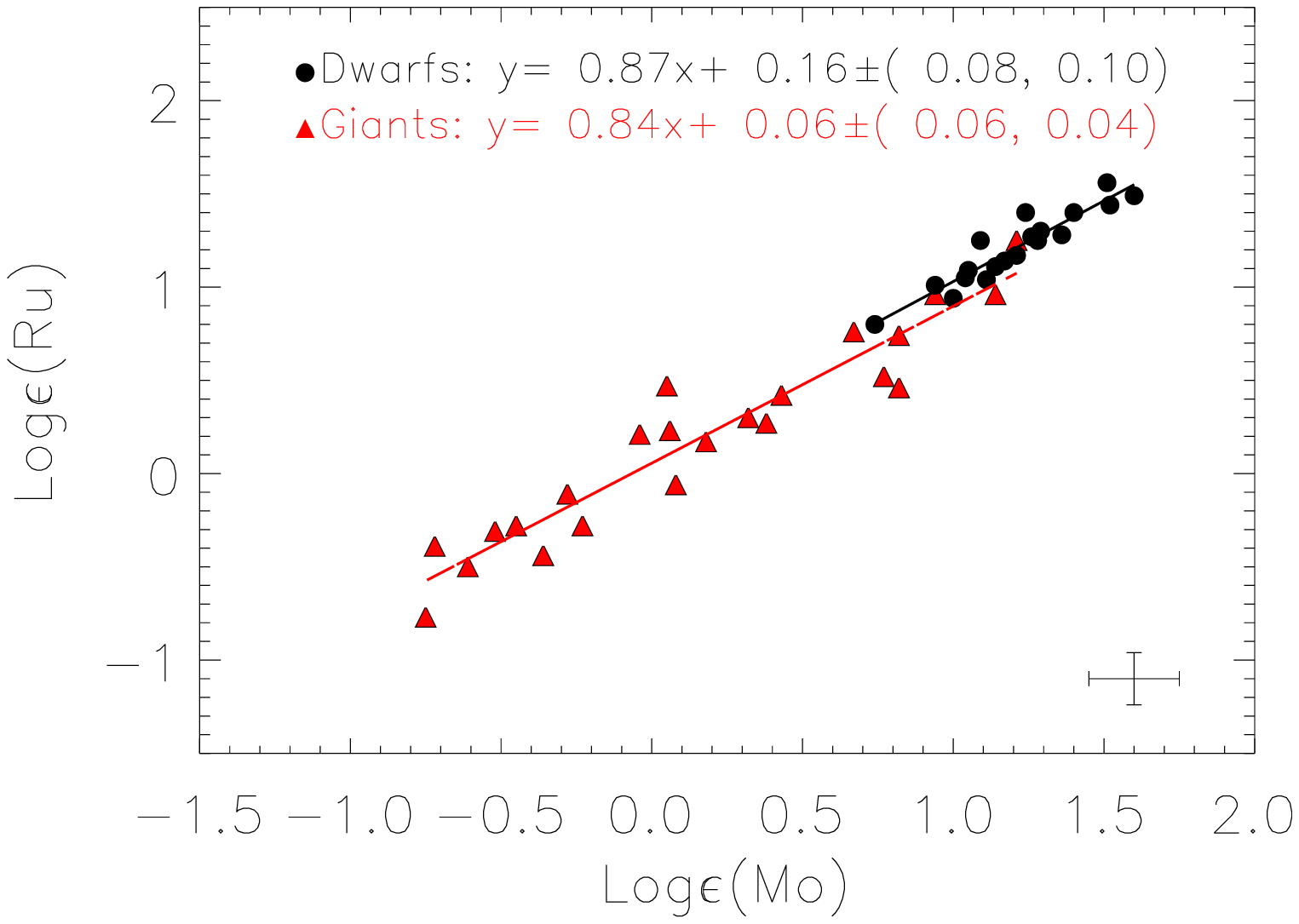}
\includegraphics[width=0.5\textwidth]{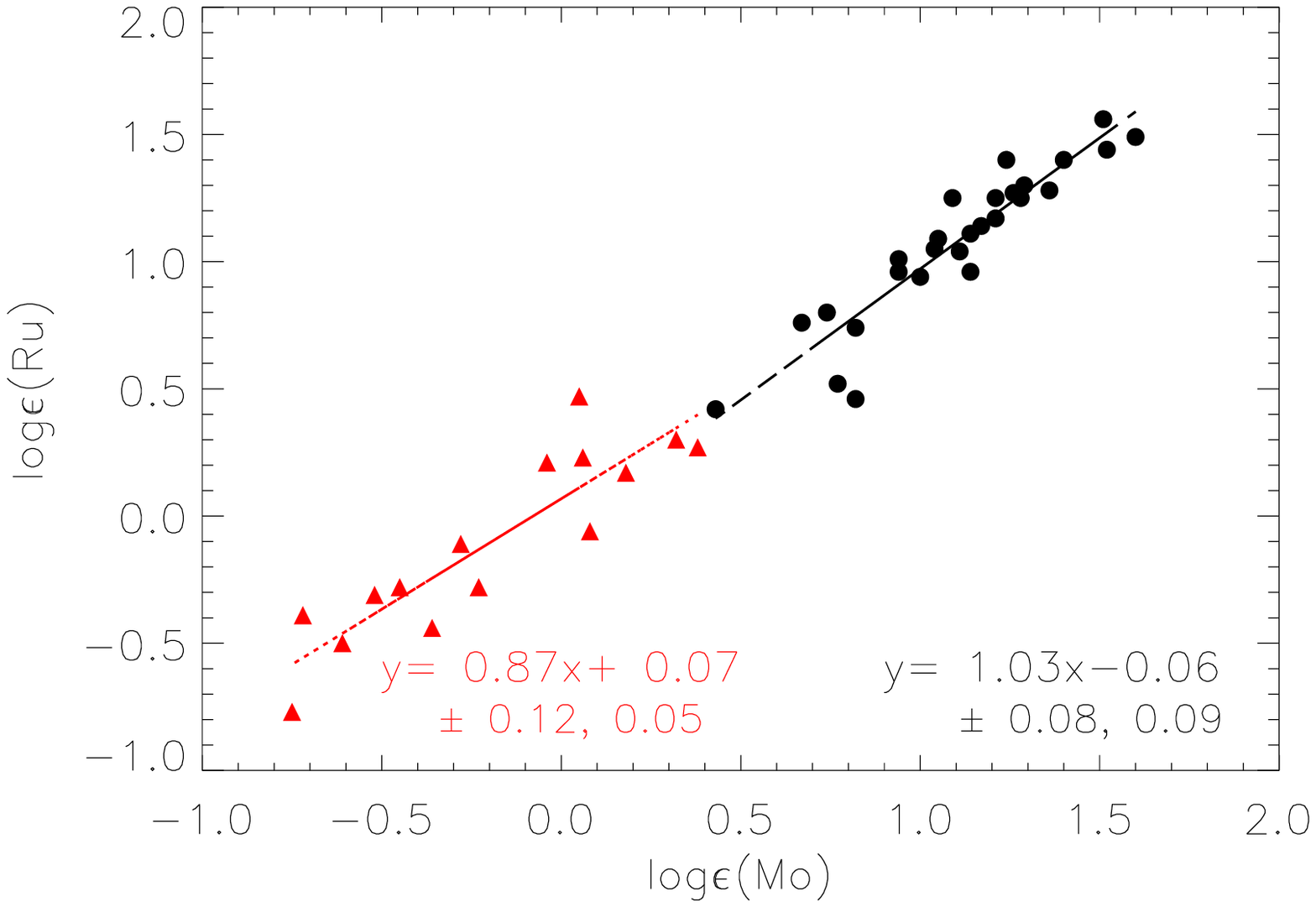}
\caption{Top: Weak correlation between Mo and Ru. Dwarfs (filled black circles) and giants (filled red triangles) with fitted lines plotted on top. Bottom: Linear trends fitted to Mo and Ru abundances in the automatically assigned clusters. Red points are generally more metal-poor than black points. \label{moru}}
\end{center}
\end{figure}

To ease the comparison made in Figs. \ref{moru} to \ref{ru2}, the formation process of each element is listed in Table \ref{processes}. The values listed are from \citet{arland}, however, in \citet{bist14} most of the s-process elements have increased by $\sim 7$\% except for Zr, for which the s-process fraction has decreased to $\sim51$\%.

\begin{table}
\caption{Formation process of Sr -- Eu with percentages from \citet{arland}.}
\begin{tabular}{ccc}
\hline
\hline
Element& Process (in \%)& Comment\\
\hline
Sr & 85\% s & Mostly weak s \\
Zr & 83\% s & Mixed: r+s+weak r \& s \\
Mo & 50\% s & Mixed: p+r+s\\ 
Ru & 32\% s & Mostly weak r\\
Pd & 46\% r & Partial weak r contribution\\
Ag & 79\% r & Predominantly weak r\\
Ba & 81\% main s & Main r at low [Fe/H] \\
Eu & 94\% main r& Always main r\\
\hline
\hline
\label{processes}
\end{tabular}
\end{table}
\subsection*{Interpreting the linear trends}
To extract the similarity in formation processes, absolute (log $\epsilon$) abundances of Mo and Ru are compared to other trace elements. If the two compared elements are formed by the same process, we expect to find a 1:1 correlation; that is, the fitted line should have a slope of 1.0. One indication of several, competing formation processes is a larger star-to-star scatter. This will be expressed as a larger uncertainty on the fitted slopes.
The trends (lines) have been fitted using a linear least-squares method that, when tested, turned out to be very robust. The least-squares method was tested against a robust least absolute deviation method and a minimum $\chi^2$ fit. All yielded the same results within the uncertainty of the fit. However, when removing a few stars from the sample, the linear least-squares method turned out to be more consistent in yielding robust slopes. Upper limits have been removed to ensure a cleaner trend of the two elements compared. Since the comparison has been carried out on absolute abundances, no obscuration from other elements such as Fe or the choice of solar abundance have been introduced. Owing to the small abundance uncertainties ($\sim 0.15$\,dex), we can require very tight correlations between the two elements. 
Furthermore, to accept the correlation, we also require that the star-to-star scatter is low. This is quantified and constrained through a 1$\sigma$ uncertainty of the fitted lines being similar to or less than the abundance uncertainty of 0.15\,dex. 
However, if the gas has been mixed or diluted with contributions from other formation processes, the correlation is  {\it weaker}, 
and the slope will deviate from 1.0 by $\sim 0.15$ or more. 
Phrased differently, if the slope is below 0.85 or above 1.15, one of the two elements must be enriched by a different secondary process, which will break the 1:1 correlation. 

Two different fitting approaches have been followed: 1) We fit dwarfs and giants separately, since we expect them to be affected differently by NLTE effects, and 2) we carry out an automated cluster analysis, where different weights are assigned to the data, which ensures that the data with similar properties are placed in a cluster.  

\begin{table}
\caption{Properties, number of members, and median metallicity of automatically optimised clusters of data.}
\begin{tabular}{ccc|cc}
\hline
\hline
Elements & \multicolumn{2}{c}{`metal-poor cluster'} & \multicolumn{2}{c}{`metal-rich cluster'}\\
       & members & [Fe/H] & members & [Fe/H]\\
\hline
Mo -- Sr & 15 & $-1.83$ & 27 & $-1.20$\\
Mo -- Zr & 16 & $-1.64$ & 26 & $-1.20$ \\
Mo -- Ru & 15 & $-1.83$ & 27 & $-1.20$ \\
Mo -- Pd & 14 & $-1.58$ & 27 & $-1.19$ \\
Mo -- Ag & 15 & $-1.64$ & 26 & $-1.20$ \\
Mo -- Ba & 7 & $-2.09$ & 34 & $-1.20$ \\
Mo -- Eu & 13 & $-1.99$ & 29 &$-1.25$\\
\hline
Ru -- Sr & 22 & $-2.10$ & 30 & $-1.12$ \\
Ru -- Zr & 23 & $-1.64$ & 28 & $-1.19$ \\
Ru -- Pd & 16 & $-1.99$ & 35 & $-1.25$\\
Ru -- Ag & 18 & $-1.92$ & 32 & $-1.25$ \\
Ru -- Ba & 15 & $-2.13$ & 36 & $-1.30$ \\
Ru -- Eu & 17 & $-1.48$ & 33 & $-1.30$\\
\hline
\hline
\label{2clusters}
\end{tabular}
\end{table}

In the first case, the offsets in the slopes fitted to the dwarfs and giants can also be taken as an expression of how important NLTE effects might be for the two elements shown. In the second case, the properties of the data have been taken into account. For a given number of clusters, the centre is determined, and the data is placed in a cluster via a minimisation of the data point's distance to the centre. The clustering was done in IDL using the routines `cluster\_wts' and `cluster'. These routines uses a k-means clustering where the initial clusters are chosen randomly, and data points are moved between clusters by minimising the variability within the cluster, thereby increasing the difference between clusters.  Since random clusters are initially assigned to the data every time the routine runs, different results could be obtained, especially for scattered data points. However, this is not the case, and in the $\sim 100$ test runs, the same clusters have always been obtained for this sample.
\begin{figure*}
\begin{center}
\includegraphics[width=0.9\textwidth]{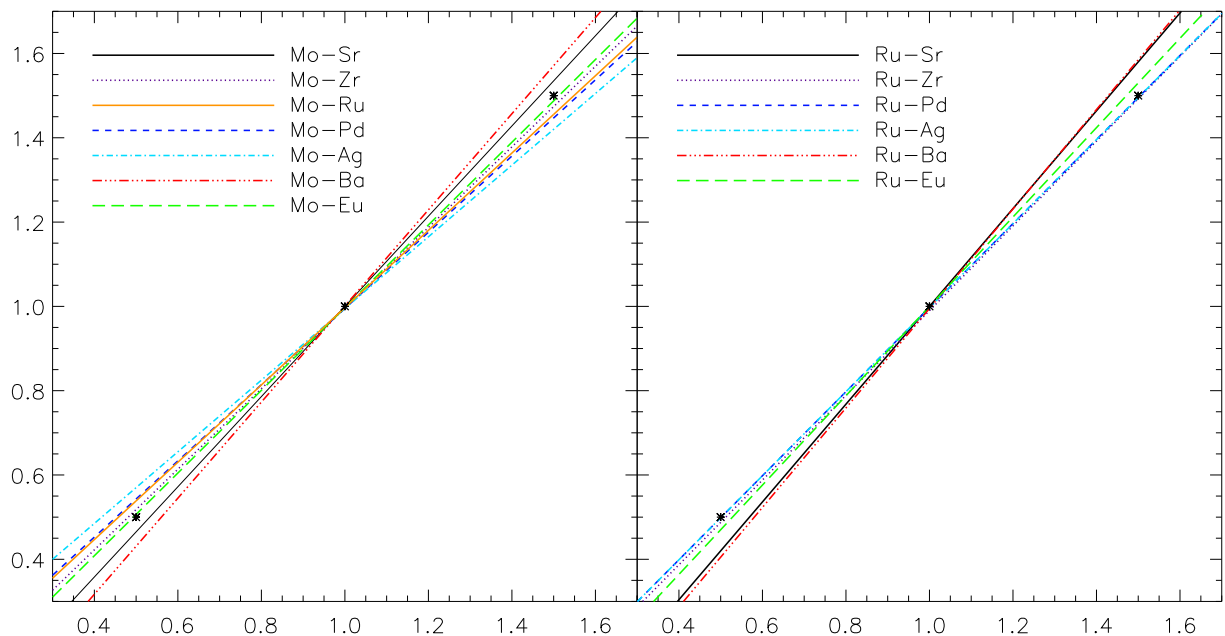}
\caption{Single trends fitted to Mo abundances vs Sr - Eu (left), and Ru vs Sr - Eu (right). Three asterisks indicate a 1:1 correlation. \label{cluster}}
\end{center}
\end{figure*}

We find that at most two clusters are needed to describe the data, otherwise the clusters get very small and are less significant with respect to a physical formation process.
The abundances can be well fitted using one or two clusters. Figure \ref{cluster} shows a summary of the correlations between Mo and Sr to Eu and between Ru and Sr to Eu. From this figure, Mo is seen to closely correlate with most of the elements, except for Ag. Thus, Mo seems to have a very mixed origin even at low metallicities. Ruthenium, on the other hand, shows the tightest correlation with Ag, Pd, and Zr; it has less in common with the production channel of Eu, and hardly anything to do with the s-processes creating Sr and Ba. In most cases, the lines fitted are associated with very low uncertainties ($<0.08$), except for the trends between Mo and Sr or Ag and Ru and Sr. 
The weaker trend between Mo and Ag plus the large uncertainty associated with the linear fit between weak r-process elements and Mo could indicate that the weak r-process might affect Sr, Mo, Ru, and Ag differently at different times or metallicities. 

We therefore explored what happens when two clusters are fitted instead of just one.
 For each of the two clusters, we calculated the median of the metallicity in the cluster. The cluster size and median metallicity are listed in Table \ref{2clusters}. In the cluster fitting case we generally find significantly different [Fe/H] median values. In most cases, the metal-poor cluster consists of metal-poor giants and dwarfs and one or two more metal-rich dwarf stars. The `metal-rich' (median [Fe/H]$>-1.5$) cluster mainly consists of more metal-rich dwarf stars with a few giants or a few metal-poor stars (2 -- 5 stars with [Fe/H] $<-2$ depending on the pair of elements).   
Assigning the data points to clusters seem to lead to a natural division of a more metal-poor sample versus a metal-richer sample, where the former most likely will be dominated by the r-process and the latter contaminated more by the s-process. Thus, by splitting the sample into sub-clusters we seem to gain more information on the underlying physical formation processes, rather than by looking at dwarfs and giants separately. The clusters provide a stronger lead on the behaviour of the formation process, and we proceed by fitting linear trends to the clusters until further notice on the actual NLTE abundance behaviour.  

The trends fitted to the giants generally agree with those of the metal-poor cluster. This makes sense, since the metal-poor cluster mainly contains metal-poor giant stars.

\subsection*{Trends from figures}
\begin{figure}
\begin{center}
\vspace{-0.3cm}
\includegraphics[width=0.49\textwidth]{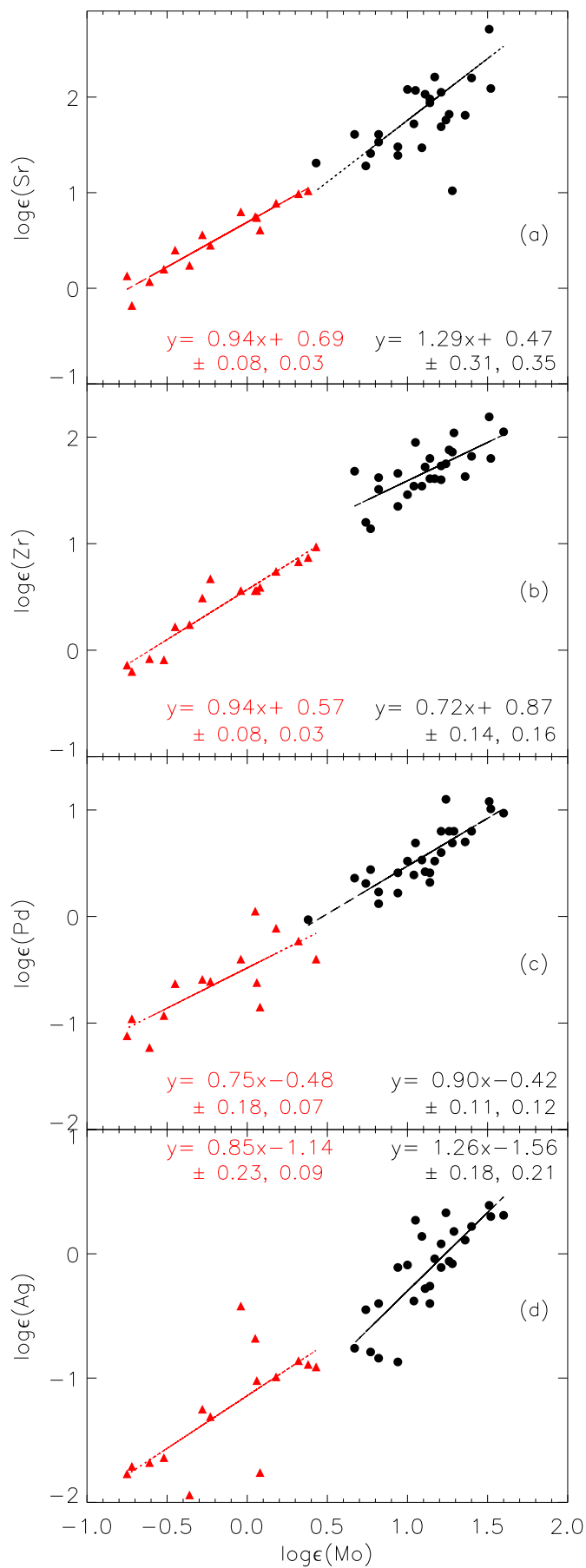}
\caption{Absolute abundance of Mo vs Sr, Zr, Pd, and Ag.}
\label{mo1}
\end{center}
\end{figure}

\begin{figure}
\begin{center}
\vspace{-0.3cm}
\includegraphics[width=0.49\textwidth]{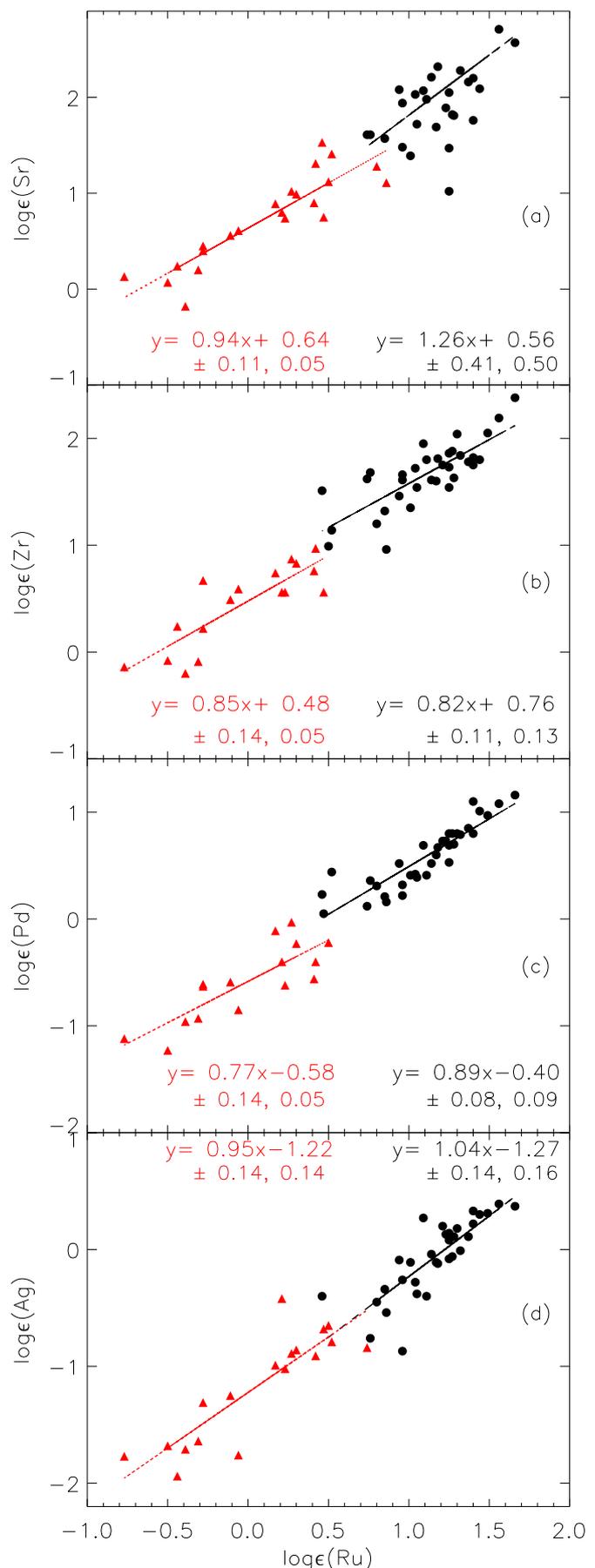}
\caption{Absolute abundance of Ru vs Sr, Zr, Pd, and Ag.}
\label{ru1}
\end{center}
\end{figure}

\begin{figure}
\begin{center}
\includegraphics[width=0.49\textwidth]{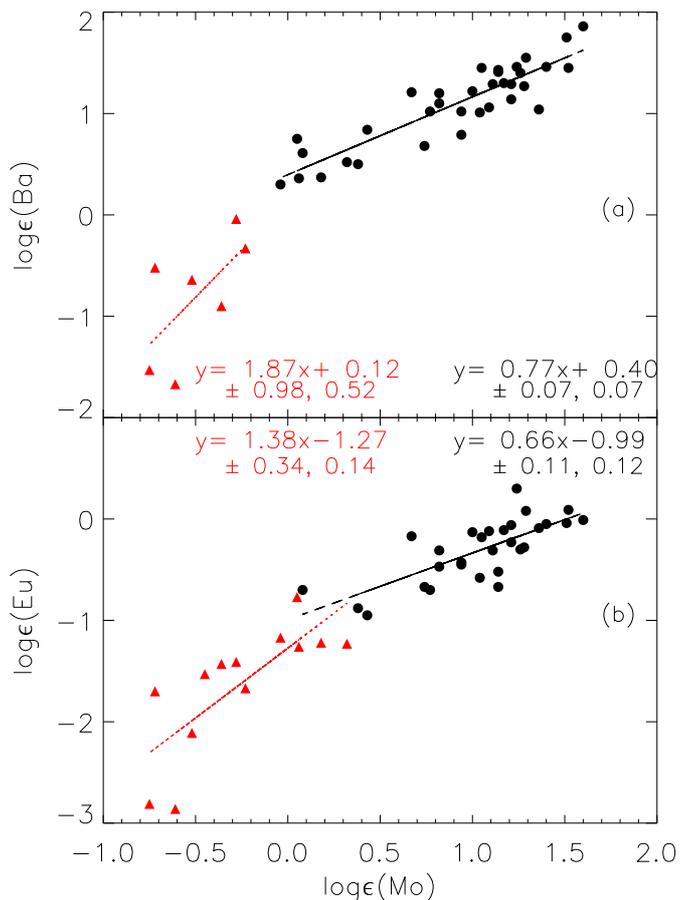}
\caption{Mo abundances compared to those of Ba and Eu. \label{mo2}}
\end{center}
\end{figure}

\begin{figure}
\begin{center}
\includegraphics[width=0.49\textwidth]{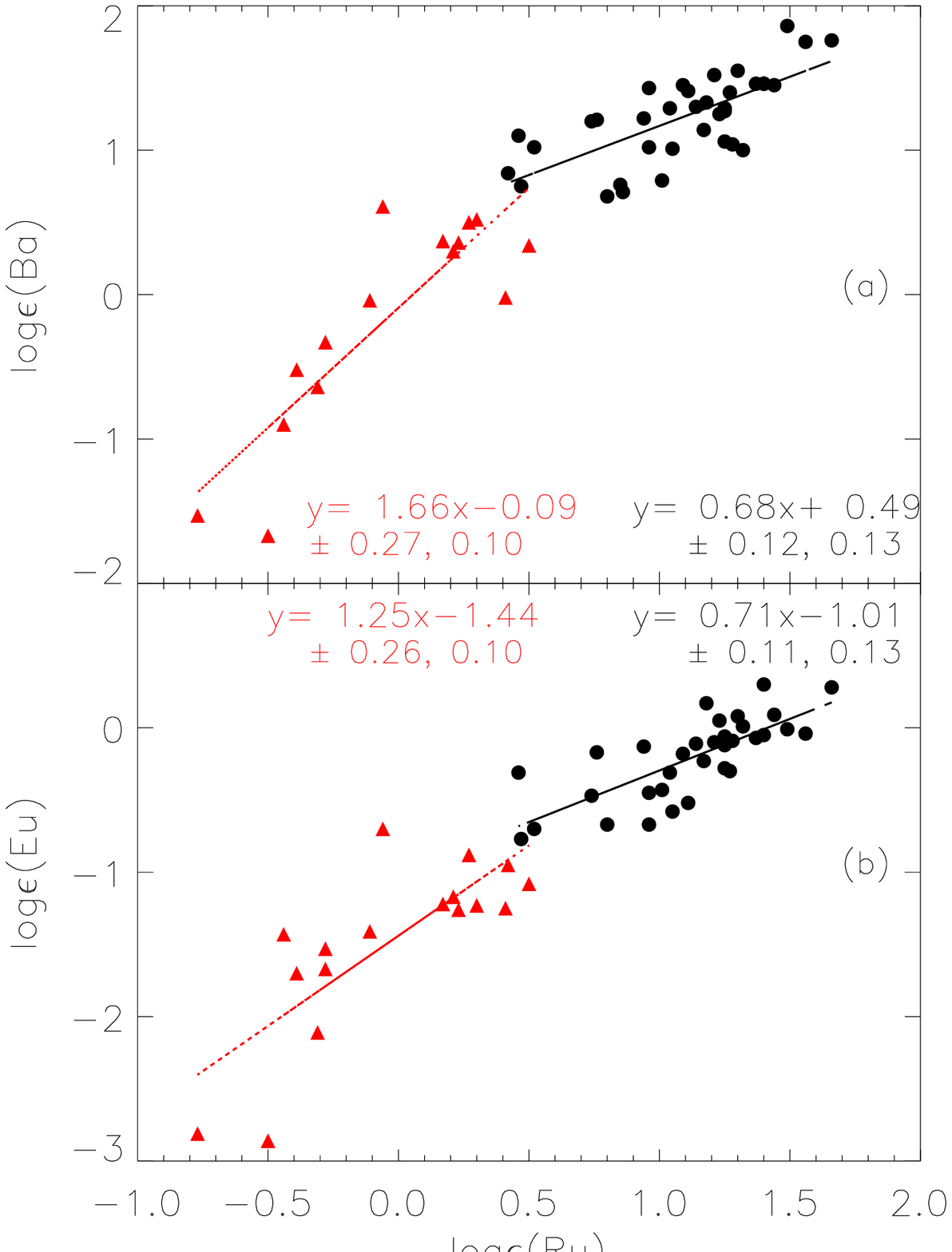}
\caption{Ru abundances compared to those of Ba and Eu. \label{ru2}}
\end{center}
\end{figure}

All the fitted lines, slopes, and intersections with the y-axis are given in each of the panels in Figs. \ref{moru} to \ref{ru2}. Immediately below these linear equations, the 1$\sigma$ uncertainty on the slope and intersection is given in parenthesis.

We start by comparing Mo and Ru in Fig. \ref{moru}, where we find correlations within the uncertainty at both higher and lower metallicity.
This is not surprising since the isotopes of these elements are created by similar processes. However, to pin down the reason for these trends and understand the influence of the weak r-process, we need to compare Mo and Ru to elements with well-determined formation processes.
In the following Figs. \ref{mo1} to \ref{ru2} the red triangles are generally the more metal-poor stars, while the black circles are more metal-rich.

From panels a) and b) in Fig. \ref{mo1} we find almost 1:1 correlations between Mo and Sr and Mo and Zr, respectively, at lower metallicity. This could indicate that weak s-process yields have been incorporated in stars with metallicities below [Fe/H] = $-1.83$. The trends are clean (a slope of 0.94$\pm0.08$) and the uncertainties low. On the other hand, from Table \ref{processes} we see that 15\% of Sr is created by a process that is different from the weak s-process. It is different from the weak r-process (see lower panels of Fig. \ref{mo1}), but it could be a sort of lighter element primary process (LEPP), such as an $\alpha$-process or a $\nu$p-process \citep{frohlich}. Another formation channel is the charged-particle process described in \citet{qian}. At low [Fe/H] either of these processes could be driving the enrichment of elements from Sr up to Mo instead of the weak s-process. At higher metallicities the slopes in the two top panels clearly deviate from unity, and the uncertainty (star-to-star scatter) is large. This could indicate that there are several formation processes creating Mo at higher [Fe/H]. One option would be the p-process or the earlier mentioned $\alpha$-/$\nu$p-process, which would explain the correlation between Mo and Ru at higher [Fe/H] since their lightest isotopes are created by a p-process.
In the following two panels (Fig. \ref{mo1} c) and d), Mo is seen to correlate with Pd at higher metallicity, indicating that both elements may receive a large contribution possibly from a weak or main s-process. However, the remaining trends in these panels do not show a 1:1 correlation, but are instead described by large star-to-star abundance scatter resulting in uncertain linear trends. Based on this, we do not believe that Mo is predominantly created by a weak r-process, but it may receive a minor contribution.

Continuing to compare Mo to Ba in Fig. \ref{mo2} a), we see that the two elements show a tight correlation at higher [Fe/H], but it clearly deviates from a line with slope 1.0. This indicates that a different process like the weak s- or the p-process interferes and creates Mo in addition to the main s-process. At low [Fe/H], the cluster is small and the data is scattered. This could also indicate that this study only includes a few main s-process dominated stars below [Fe/H] =$ -2$ or that another process is interfering. A larger sample of both metal-poor and metal-rich stars containing Mo and Ba would be needed to clarify this.
The largest amount of Mo is also not created by the main r-process, which is responsible for the production of Eu (see Fig. \ref{mo2} panel b). Molybdenum and Eu show a large scatter and uncertainty and a deviation from a 1:1 correlation when split into sub-clusters. This could, as mentioned above, indicate that even larger samples are needed. A single cluster indicates that Mo and Eu correlate almost 1:1 with a low scatter. However, differences in formation processes or traces of a possible late onset of a process (e.g. the main s-process) seem to be erased when a single cluster is enforced. This makes it hard to tell if small number statistics are playing a dominant role, or if different processes are setting in at different times. Increasing the sample of stars with Eu, Ba, and Mo in both metal-poor and metal-rich stars would help clarify this issue. 

Ruthenium is different from most weak s-process-dominated elements at higher metallicity (see Fig. \ref{ru1} panel a), b), and less so in c); cf. Table \ref{processes}). This is expressed in slopes that differ by more than 0.25 from unity and uncertainties of up to 0.41. This uncertainty is the largest derived and is found between Ru and Sr in both one and two clusters. This picture changes at lower metallicity, and if the weak s-process is already efficient at this low metallicity, these trends and differences between higher and lower metallicity become hard to explain. Again, we note that a LEPP, $\alpha$- or $\nu$p-process may be responsible for the formation of Sr - Ru, in which case the issue with the weak s-process is circumvented. 

Very direct and clean almost 1:1 trends are found between Ru and the weak r-process element Ag (Fig. \ref{ru1} d). This shows that Ru receives a dominant contribution from the weak r-process regardless of metallicity, and the influence of the weak r-process may also play a role in the correlations we find at lower metallicity between Ru and Zr, but less so between Ru and Sr. This could indicate that Zr is not mainly produced by the s-process, as predicted by \citet{arland}, but rather $\sim 50/50$ r/s, as recently indicated in \citet{bist14}.

Finally, Fig. \ref{ru2} a) shows Ru versus Ba, where the abundances of these elements do not correlate at any metallicity, and the slopes deviate the most from 1.0, so the main s-process has the lowest influence on the Ru abundances. This is seen in one and two clusters as well as in dwarfs and giants. Panel b) of the same figure shows a large star-to-star scatter and slopes different from unity. Both indicate that the main r-process is not dominating the Ru production, though it is still contributing more than the main s-process is. 

Even when considering that Ba could be produced by a main r-process at low [Fe/H] (cf. Table \ref{processes}), we still find clear differences between Mo and both heavy elements (Ba and Eu), as well as between Ru, Ba, and Eu. This shows that Mo and Ru are not predominantly produced by the main r-process and that the process driving their correlations may be a sort of LEPP. This process is clearly different from the main r-process, which is responsible for the formation and slopes of Eu, and also for Ba at low [Fe/H].

\begin{figure*}
\begin{center}
\includegraphics[width=0.9\textwidth]{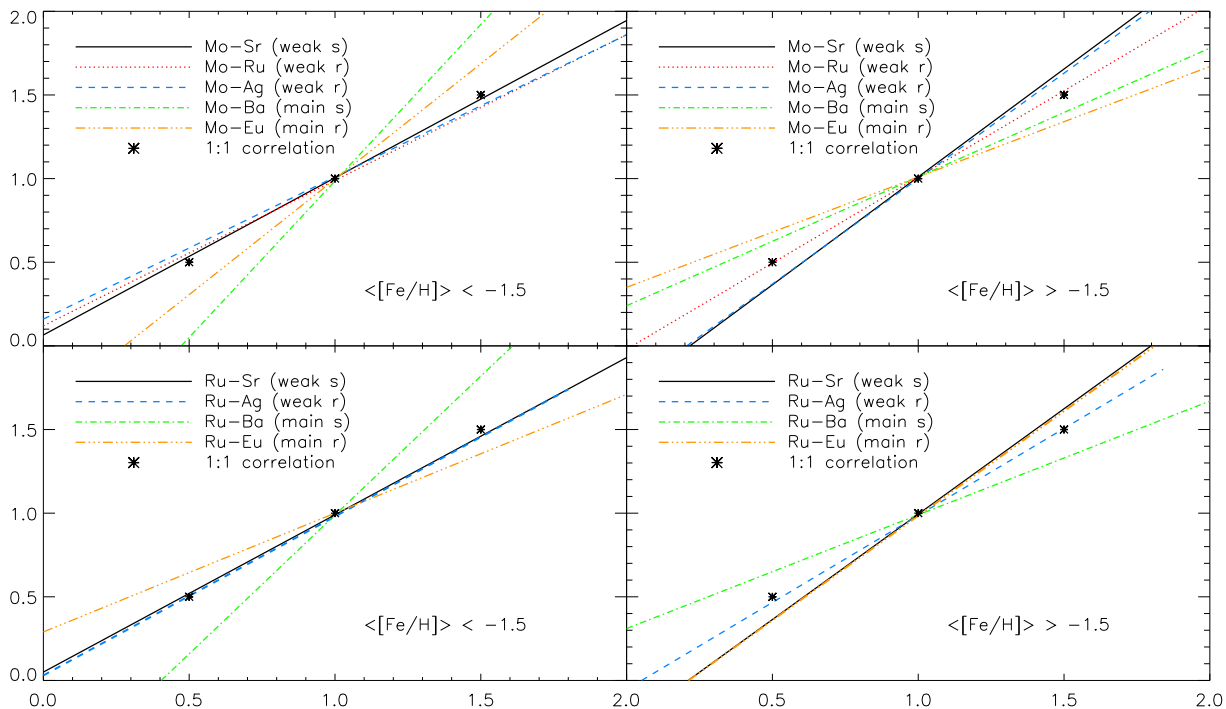}
\caption{Comparison of slopes fitted to the elements given in the legend. Top panels show Mo with respect to other key elements (Sr--Eu), while the bottom panels show Ru. Median metallicity of the cluster is stated in each figure.
\label{multi}}
\end{center}
\end{figure*}

To ease comparison of the cluster slopes, they have been summarised in Fig. \ref{multi}. These have been shifted for the sake of clarity to go through the point (1,1), but all the slopes have been preserved. This also allows a very direct comparison to the slopes obtained from only one cluster with one line fitted to the full sample (see Fig. \ref{cluster}).    


With the adopted log gf values, the focus on continuum placement and blends, the elemental abundances are as precise as currently possible, and we consider the trends trustworthy. We are aware that larger sample sizes will most likely improve the trends and make the according results more robust.

Generally, at high [Fe/H], Mo is seen to correlate with Ru and Pd, which in the case of Ru could mean a considerable p-process contribution\footnote{The p-process may be metallicity sensitive and show a secondary nature according to \citet{pigna13}}. At lower [Fe/H] Mo correlates directly with Sr and Zr. This could point towards early weak s-process contributions in which case the weak s-process range goes up to and includes Mo but not Pd; however, this may also be true for the $\alpha$- or $\nu$p-process, for instance. If the weak s-process is the source, it would strengthen the scenario described by \citet{pigna13}. In one case, \citet{pigna13} show that Mo and Ru would require a very high s-process rate leading to a disagreement between weak s- and main s-process, while the p-isotopes of Mo and Ru can be reproduced via a `cs-process' in massive stars (M$\sim25M_{\odot}$) or via a $\nu$p-process \citep{frohlich}.
Ruthenium is seen to be a less convolved element than Mo, and Ru is mainly formed by the weak r-process as is Ag.

The study by \citet{lorusso_rp} combining experiments and theory shows that the formation process of $^{96}$Ru is unlikely to be an rp-process.  The exact environment and sites of these processes are far from resolved, and we leave further discussions to a future paper (see e.g. Fr\"ohlich et al, 2014 in prep).

\section{Discussion \label{discus}}
Even with elemental abundances we are capable of distinguishing between the formation processes and find the one that donate the dominant amount to each of the elements. However, considering isotopic abundances of presolar grains allows a much more detailed and direct comparison between the isotopic abundance and associated formation process.

\subsection*{Presolar grains}
The presolar grain abundances are taken from the presolar database \citep{presolar} at St. Louis University\footnote{http://presolar.wustl.edu/pdg and E. Zinner priv. comm.}. We compare abundances of the SiC grains within the database that have known abundances of Mo and minimum one of the other heavy elements shown in Sect. \ref{correl}. Presolar silicon carbide grains are the best studied presolar mineral phase. This is due to the relatively high abundance in primitive meteorites and to the size of the SiC grains, which allows isotopic analyses, of the major and of many trace elements, of individual grains. The presolar SiC grains are categorised based on the abundant elements C and Si, as well as on the trace elements that are present within the minerals. Isotopic data exist for N, Mg, Ca, Ti, the noble gases, and heavy refractory elements (e.g. Zr, Mo, La, Nd, Sm, Ru, and Ba). Based on the isotopic composition of C, N, Si, and the abundance of radiogenic $^{26}$Mg, six different populations of SiC grains are discerned: the mainstream grains, which make up the majority of the grains (around 90\% of the total), and the minor types A, B, X, Y, and Z. For our study, the X grains are of particular interest because their C- and Si-isotopic compositions can be explained by mixing of matter from the C- and Si-rich zones in a type II SN.
The X-grains are characterised by enrichments (relative to solar isotopic abundance) in $^{12}$C (most grains), $^{15}$N, and $^{28}$Si. Many of the grains have isotopic over-abundance in $^{26}$Mg \citep{Hoppe94,Virag92,Zinner91}, $^{44}$Ca \citep{Nittler96}, and $^{49}$Ti \citep{Hoppe00,Amari00}, which likely came from the radioactive decay of $^{26}$Al (t$_{1/2}\sim 0.7$ Myr), $^{44}$Ti (t$_{1/2}\sim60$ yr), and $^{49}$V (t$_{1/2}\sim330$ d) after grain formation.  We refer to \citet{lodders05}, and \citet{hoppeetal10} for further details on presolar grains.

The abundances of presolar grains are given as ratios on the $\delta$-abundance scale between the meteoritic grain and a standard, 
\begin{equation}
\delta^iX =\left(\frac{(^iX/^jX)_{grain}}{(^iX/^jX)_{standard}} -1\right)\cdot 1000,
\label{delta}
\end{equation}
where X is the element, and i and j different isotopes. The standard is often taken from a terrestrial sample or the reference values from \citet{anders}.
The total absolute meteoritic abundance cannot be extracted from these measurements, so we are forced to use relative isotopic abundances, which among other things can describe the fraction of r-/s-process present in the grain. Assuming the natural ratios of p-, r-, and s-isotopes are universal, this ratio should be descriptive of both the grain and the total meteorite, as well as the stars. Furthermore, neither Zr nor Ba are siderophile elements (Mo only slightly so), nor are they strongly separated between the gas and solid phases. 

Recently, \citet{Mann_Rupart} have shown that Ru was not partitioned, not even under high pressure or temperature. Both Mo and Zr are refractory elements and even though Ba is less refractory, it is still far from being volatile. It therefore seems most natural that their abundances are representative of the original gas composition. 
Actually, \citet{lodders2010} investigated how the siderophile nature affected $\log \epsilon$ Mo abundances in the Sun, now and just after the formation of the Sun. The difference can be found by comparing the solar Mo in their Tables 3 and 6, which shows that the difference is 0.07\,dex and no difference for Ru. However, here we have chosen not to correct for these changes in Mo or other elements, since we have a mixture of dwarfs and giants, and they span a broad range of ages. The giants have a convective atmosphere, and we do not know how this would affect these estimates. Furthermore, we also have stars that are older than the Sun. Combining the variation in ages and convection, this would lead to a difference in the solar reference value of Mo. In either case this value will most likely remain small and stay below the adopted uncertainty. Therefore, it will not have any impact on the correlation trends shown here, and we find it safe to discard this.

We now focus on Zr, Mo, and Ba because Zr and Ba seem to be some of the most studied elements in combination with Mo. The most abundant isotope is used as the reference in the presolar database, and the `j' isotope in Eqs. \ref{delta} and \ref{logE}, is $^{94}$Zr (s-dominated), $^{96}$Mo (s-only), and $^{136}$Ba (s-only) when we take the data from the presolar database. This means that we have a pure s (or at least s dominated) denominator in the relative log $\epsilon$-ratio in Eq. (\ref{logE}).

\begin{figure}
\begin{center}
\vspace{-0.3cm}
\includegraphics[width=0.49\textwidth]{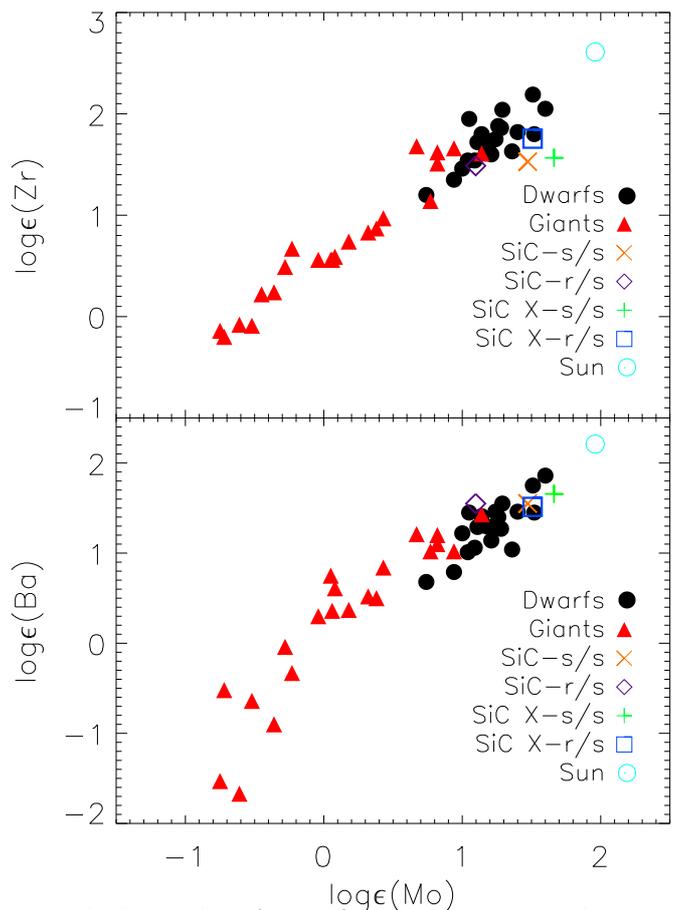}
\caption{Comparing r/s and s/s ratios in presolar grains to stars from our samples including the Sun as reference.}
\label{metfig}
\end{center}
\end{figure}

To obtain the relative ratios, we first calculate the pure grain ratio of $\frac{^iX}{^jX}_{grain}$ by removing the Earth reference from the $\delta$-ratio in Eq. \ref{delta}. The numerator isotopes `i' for Zr are 91, the most s-process-dominated Zr isotope, and 96 (r-only). Similarly we select r- and s-only isotopes for Mo and Ba to calculate r/s and s/s ratios. For Mo the two numerator isotopes are 98 (second most s-dominated isotope after 96) and 100 (r-only), while for Ba isotope 135 (most r-dominated Ba isotope existing) and 138 (second most s-dominated isotope after 136) were chosen. The choice of isotopes was made in accordance with Table 1 in \citet{chrisrev}. The log $\epsilon$-ratios are given by Eq. \ref{logE} and plotted on top of the stellar abundances in Fig. \ref{metfig}:  
\begin{equation}
\label{logE}
\log \epsilon = \log \left(\frac{\frac{^iX}{^jX}_{grain}}{\frac{^iX}{^jX}_{A\&G}}\right) + 1.554.
\end{equation}
Here we selected the meteoritic abundances (A\&G) from \citet{anders} to ensure a consistent reference scale. The constant (1.554) is a conversion factor taken from \citet{anders}, which converts the abundances from a meteoritic $10^6$ Si atom scale to an astrophysical $10^{12}$ H atom scale.
The $\delta$-abundances of SiC are originally from \citet{nic97} and the SiC X from \citet{Pellin06}.   

In the top panel of Fig. \ref{metfig} we have shown the total stellar abundance of Mo, Zr, as well as the average of the fractions of r- and s-processed material derived from presolar grains. 
The r/s and s/s ratios are seen to agree well with the total mixed stellar abundance derived from dwarfs with a metallicity of [Fe/H] $=-1.5$ to $-1.1$. We have also included the Sun (abundances from \citealt{anders}) in Fig. \ref{metfig} to guide the following discussion.

The presolar SiC grain r/s ratios match the abundance of the metal-poor dwarfs down to [Fe/H]$\sim -1.5$. Above [Fe/H]$=-1.5$, the r/s ratios must therefore be the same everywhere in the Galactic ISM, and the gas composition thus seems homogeneous. The smooth decreasing trend down to 0.0 in log $\epsilon$(Mo) could be taken as an expression of this, while a deviation from the trend is seen below 0.0 (Fig. \ref{metfig} bottom panel). It could indicate that heterogeneity is mainly found below log $\epsilon$(Mo) $\sim 0$, which for this sample corresponds to [Fe/H$]<-2.3$.  Therefore we cannot support a heterogeneous nebula at these high metallicities $\sim -1.1$ to $-1.5$ (as in \citealt{Dauphas04} and \citealt{burkhardt}). The suggested decoupling between p- and r-processes (or just between r-processes; \citealt{chen04}) is more likely based on our stellar abundance correlations in Sect. \ref{correl}. 

The SiC X grains (open square) are seen to have a slightly larger r/s fraction than the mainstream SiC (open diamond see Fig. \ref{metfig}). This agrees with SiC X grains being enriched by supernovae type II \citep{hoppe96}. That the s-dominated Mo ratios are greater than the corresponding r/s ratios in mainstream SiC grains shows that these SiC grains are depleted in r-process, which agrees with \citet{Pellin06} and \citet{nic97}. In the mainstream SiC grains, the isotopic s-abundances are larger than the stellar abundances. This also agrees with the facts that SiC grains are generally thought to be directly enriched by an AGB stars and that the s-process created isotopes might be formed at the expense of the r-isotopes. Since the grains are more s-process-enriched than the stars, it could indicate that the gases in stars are more diluted or differently mixed than the SiC grains. The mixing processes need to be better constrained, both within the AGB stars, which directly affects the SiC 
grains, and in the ISM where yields from SN and 
AGB will be incorporated into stars. Until the mixing is better understood, stronger conclusions cannot be drawn based on either elemental or isotopic abundances.
The meteorites seem to have formed from a gas similar to that of the stars at [Fe/H]$=-1.5 $ to $-1.1$. This agrees with the presolar grains being made prior to (at a lower metallicity than) the Sun. 

From the bottom panel of Fig. \ref{metfig}, both SiC grains are seen to contain more s-process- than r-process-created material. This is in good agreement with what we found in the top panel. Furthermore, the abundance compositions follow a Galactic chemical evolution scheme, where the mixture of supernova r- and  AGB star s-process material enrich the dwarfs and giants around and above $-1.5$ in [Fe/H], and yet stay below the solar ratios. 
In general, the presolar grains are slightly more enhanced in s-process material than the metal-poor dwarfs around [Fe/H] $=-1.5$ to $-1.1$. Despite Ba being a less refractory element than Mo, and Mo more siderophile than Ba, the Ba/Mo ratio from the grains seems to match that of the stars\footnote{This could indicate that the corrections from \citet{lodders2010} are indeed small or cancel out.}. This could help put constraints on how efficient the mixing processes are in mixing AGB and/or SN gases into later generations of stars. 

This comparison of stars and grains show that both seem to share the same mixture of r/s-gases and therefore also formation processes around [Fe/H] $\sim -1.5$ and above. Figure \ref{metfig} indicates that the stars around this metallicity show a mixture of both r- and s-processes, which means that s-process traces are seen down to (and likely below) [Fe/H]$=-1.5$. Below this metallicity \citet{roed10} found little or no s-process in their sample stars, and \citet{ruth11} conclude that s-process yields `were largely absent' in their sample. This is in concordance with \citet{trav04} where the weak s-process contribution to halo stars is found to negligible. Our comparison to grains indicate that this cut in metallicity may be moved down in metallicity to somewhere between [Fe/H]$=-1.5$ and $-2.5$. The lower value we base on the increasing abundance star-to-star scatter found in \citealt{hansen12} at this metallicity. Furthermore, this value is in good agreement with the average metallicity ($-2.58$) we find from the stars below log $\epsilon$(Mo) = 0 in Fig. \ref{metfig}.

\section{Summary and conclusion \label{concl}}
We summarise the outcome from the stellar abundances first, and then conclude on both stellar and meteoritic abundances.
Ruthenium is in all cases seen to correlate almost perfectly with silver, and this provides a strong observational indication of ruthenium being created by the weak r-process. The weak r-process is dominating the formation of Ru. Smaller amounts of Ru are created by the p- and r-processes, as well as the weak s-process. The main s-process is in this connection the poorest donor.

From Fig. \ref{multi}, Mo is seen to have more in common with the lighter elements than the heavy elements. Based on these observations, Mo can be considered as a highly mixed element, where contributions from p-, main, and weak s-processes are all mixed with smaller contributions from the main r-process. 
The influence of the weak r-process is of less importance. The weak s-process seems to be able to form Mo. This could confirm the extension of the weak s-process to include Mo, as found by \citet{pigna,pigna13}, despite the mentioned model complications. The exact onset of the weak s-process is hard to trace. 

According to \citet{trav04} the weak s-process contributes little to the metal-poor halo stars owing to the metallicity dependence of the process, and they point towards a LEPP origin for Sr--Zr instead of the s-processes. At low [Fe/H] the lighter elements studied here may very well be produced by a $\nu$p-, $\alpha$-, charged-particle, or some other primary process rather than the weak s-process. In either case, this process should be different from the weak r-process creating Ru -- Ag.

To understand the formation of Mo and Ru in greater detail at low metallicity, the metallicity dependence of p-, s-, cs-processes, etc. need to be understood first. Moreover, the mass of the production site is another important quantity to constrain, in addition to the metallicity, and the combined behaviour of yields need to be explored in Galactic chemical evolution models.

By studying the two clusters instead of one, the abundance scatter and uncertainty in the fitted line between Mo and Sr is found at higher metallicity, while between Mo and Ag the scatter is found at lower metallicity. This could indicate that different processes are dominating the Mo production as a function of time or metallicity (cf. Sect. \ref{res}). To pin down the subtleties of the formation processes at this level, isotopic abundances, or even larger samples spanning broad Fe, Ru, and Mo abundance ranges are needed.

The r-/s-process mixture in presolar grains agree well with the chemical composition of the dwarf stars around [Fe/H]$=-1.5$ to $-1.1$. This could indicate that both grains and stars around and above [Fe/H]$=-1.5$ are mixed well (as also seen from Galactic chemical evolution trends), and we can therefore not support a heterogeneous presolar nebula. An inhomogeneous ISM is only expected at lower metallicities. A possibility for abundance anomalies or differing r/s or s/s ratios therefore ought to arise from differences in nucleosynthetic origin. Another possibility that might explain anomalies could be related to measuring techniques, or problems with e.g. fractionation. 
A main s-process yield of Mo and Ba is seen in stars with [Fe/H]$=-1.5$.
The difference between the total stellar abundance and the s-only isotopes from the grains could indicate that AGB yields are less efficiently mixed into stars than presolar grains. However, we need a better mixing description and hyperfine splitting for Mo to derive isotopic stellar Mo abundances to accurately probe this.

\begin{acknowledgements}
This work was supported by Sonderforschungsbereich SFB 881 "The Milky Way System" (subproject A5) of the German Research Foundation (DFG). The Dark Cosmology Centre is funded by the Danish National Research Foundation. We would like to thank the anonymous referee for constructive comments.
CJH thanks U. G. J\o rgensen, H.-P. Gail, and C. Fr\"ohlich for discussion. CJH also thanks L. Nittler, and E. Zinner for guidance and access to the presolar database.
This research has made use of NASA's Astrophysics Data System, the SIMBAD database, operated at the CDS, Strasbourg, France, and the Two Micron All Sky Survey, which is a joint project of the University of Massachusetts and the Infrared Processing and Analysis Center/California Institute of Technology, funded by the National Aeronautics and Space Administration and the National Science Foundation. 

\end{acknowledgements}

\bibliographystyle{aa}
\bibliography{MoRucorV2ny2}

\end{document}